\documentclass[twocolumn]{aastex6}
\usepackage{amsmath}
\usepackage[figuresright]{rotating}
\usepackage{multirow}
\usepackage{setspace}

\newcommand{\scl}	{\Sigma_{\rm cl}}
\newcommand{\gcm}	{{\rm g}\,{\rm cm}^{-2}}
\newcommand{\chisq}     {\chi^2}
\newcommand{\mc}        {M_c}
\newcommand{\ms}        {m_*}
\newcommand{\inc}       {\theta_\mathrm{view}}
\newcommand{\av}        {A_V}

\newcommand{\fnuobs}    {F_{\nu,\mathrm{obs}}}
\newcommand{\fnumod}    {F_{\nu,\mathrm{mod}}}

\newcommand{\fnufit}    {F_{\nu,\mathrm{fit}}}
\newcommand{\sigmau}     {\sigma_u}
\newcommand{\sigmal}     {\sigma_l}

\slugcomment{Accepted by ApJ}

\shorttitle{The SOMA Survey. I. Overview and First Results}
\shortauthors{De Buizer et al.}

\begin{document}

\title{The SOFIA Massive (SOMA) Star Formation Survey. I. Overview and First Results}

\author{James M. De Buizer\altaffilmark{1}, Mengyao Liu\altaffilmark{2}, Jonathan C. Tan\altaffilmark{2,3,4}, Yichen Zhang\altaffilmark{5,6}, Maria T. Beltr\'{a}n\altaffilmark{7}, Ralph Shuping\altaffilmark{1}, Jan E. Staff\altaffilmark{2,8}, Kei E. I. Tanaka\altaffilmark{2}, Barbara Whitney\altaffilmark{9}}
\affil{$^1$SOFIA-USRA, NASA Ames Research Center, MS 232-12, Moffett Field, CA 94035, USA\\
$^2$Department of Astronomy, University of Florida, Gainesville, FL 32611, USA\\
$^3$Department of Physics, University of Florida, Gainesville, FL 32611, USA\\
$^4$National Astronomical Observatory, Mitaka, Tokyo 181-8588, Japan\\
$^5$Departamento de Astronom\'{\i}a, Universidad de Chile, Casilla 36-D, Santiago, Chile\\
$^6$The Institute of Physical and Chemical Research (RIKEN), Hirosawa 2-1, Wako-shi, Saitama 351-0198, Japan\\
$^7$INAF-Osservatorio Astrofisico di Arcetri, Largo E. Fermi 5, I-50125 Firenze, Italy\\
$^8$College of Science and Math, University of Virgin Islands, St. Thomas, Virgin Islands 00802, USA\\
$^9$Department of Astronomy, University of Wisconsin-Madison, 475 N. Charter St, Madison, WI 53706, USA}

\begin{abstract}
We present an overview and first results of the Stratospheric
Observatory For Infrared Astronomy Massive (SOMA) Star Formation
Survey, which is using the FORCAST instrument to image massive
protostars from $\sim10$--$40\:\rm{\mu}\rm{m}$. These wavelengths
trace thermal emission from warm dust, which in Core Accretion models
mainly emerges from the inner regions of protostellar outflow
cavities. Dust in dense core envelopes also imprints characteristic
extinction patterns at these wavelengths, causing intensity peaks to
shift along the outflow axis and profiles to become more symmetric at
longer wavelengths. We present observational results for the first
eight protostars in the survey, i.e., multiwavelength images,
including some ancillary ground-based MIR observations and archival
{\it{Spitzer}} and {\it{Herschel}} data. These images generally show
extended MIR/FIR emission along directions consistent with those of
known outflows and with shorter wavelength peak flux positions
displaced from the protostar along the blueshifted, near-facing
sides, thus confirming qualitative predictions of Core Accretion
models. We then compile spectral energy distributions and use these to
derive protostellar properties by fitting theoretical radiative
transfer models. Zhang \& Tan models, based on the Turbulent Core
Model of McKee \& Tan, imply the sources have protostellar masses
$m_*\sim10$--50$\:M_\odot$ accreting at
$\sim10^{-4}$--$10^{-3}\:M_\odot\:{\rm{yr}}^{-1}$ inside cores of
initial masses $M_c\sim30$--500$\:M_\odot$ embedded in clumps with
mass surface densities
$\Sigma_{\rm{cl}}\sim0.1$--3$\:{\rm{g\:cm}^{-2}}$. Fitting Robitaille
et al. models typically leads to slightly higher protostellar masses,
but with disk accretion rates $\sim100\times$ smaller. We discuss
reasons for these differences and overall implications of these first
survey results for massive star formation theories.
\end{abstract}


\section{Introduction}

The enormous radiative and mechanical luminosities of massive stars
impact a vast range of scales and processes, from reionization of the
universe, to galaxy evolution, to regulation of the interstellar
medium, to formation of star clusters, and even to formation of
planets around stars in such clusters. 
Furthermore, synthesis and dispersal of heavy elements by massive
stars play key roles in the chemical evolution of the cosmos. In spite
of this importance, there is still no consensus on the basic formation
mechanism of massive stars. Theories range from Core Accretion models,
i.e., scaled-up versions of low-mass star formation (e.g., the
Turbulent Core Model of McKee \& Tan 2002; 2003, hereafter MT03), to
Competitive Accretion models at the crowded centers of forming star
clusters (Bonnell et al. 2001; Wang et al. 2010), to Protostellar
Collisions (Bonnell et al. 1998; Bally \& Zinnecker 2005). 

This confusion is due in part to the typically large distances
($\gtrsim 1\:$kpc) and extinctions to massive protostars (see, e.g.,
Tan et al. 2014 for a review). Massive stars are observed to form in
dense gas clumps with mass surface densities of $\Sigma_{\rm
  cl}\sim1\:{\rm g\:cm^{-2}}$ (i.e., $A_V\sim200$~mag; $A_{\rm 8\mu
  m}\sim8$~mag; $A_{\rm 37\mu m}\sim 3$~mag; adopting the opacities of
the moderately coagulated thin ice mantle dust model of Ossenkopf \&
Henning 1994). If massive cores are in approximate pressure and virial
equilibrium with this clump (MT03), then such a core with mass $M_c$
has radius $R_c=0.057 (\Sigma_{\rm cl} / {\rm
  g\:cm^{-2}})^{-1/2}(M_c/60\:M_\odot)^{1/2}$~pc. If the degree of
rotational support is similar to that in low-mass cores, then the disk
size should be $\sim100$--$10^3\:$au in radius (i.e.,
$\lesssim1\arcsec$ in size for sources at distances of
$\gtrsim1\:$kpc). The accretion rate is expected to be a
few~$\times10^{-4}\:M_\odot\:{\rm yr}^{-1}$.

Collimated bipolar outflows are observed from massive protostars
(e.g., Beuther et al. 2002) and massive early-stage cores (Tan et
al. 2016). These are thought to be accretion powered, driven by
rotating magnetic fields that are coupled to the accretion disk and/or
the protostar leading to disk winds (e.g., K\"onigl \& Pudritz 2000)
or X-winds (Shu et al. 2000), respectively. Such protostellar outflows
are expected to limit the star formation efficiency from a core to
$\sim0.5$ (Matzner \& McKee 2000; Zhang, Tan \& Hosokawa 2014,
hereafter ZTH14; Kuiper, Yorke \& Turner 2015), since they expel core
material from polar directions.

Creation of low-density outflow cavities is expected to have a
profound effect on the mid-infrared (MIR) appearance of massive
protostars (De Buizer 2006). Radiative transfer calculations of the
MT03 Core Accretion model of massive protostars (a single protostar in
an individual core) have confirmed the importance of these cavities on
the MIR to FIR images and spectral energy distributions (SEDs; Zhang
\& Tan 2011; Zhang, Tan \& McKee 2013, hereafter ZTM13; ZTH14;
Y. Zhang \& J. C. Tan 2017, in preparation). Shorter wavelength light
tends to emerge along the outflow cavity that is directed toward our
line of sight, i.e., the near-facing, blueshifted side of the
outflow. At near-infrared (NIR) wavelengths, the appearance of the
protostars is typically dominated by scattered light escaping from the
cavities. Moving to MIR wavelengths, especially $\gtrsim10\:{\rm\mu
  m}$, thermal emission from warm dust in the outflow and outflow
cavity walls makes the dominant contribution. The far-facing outflow
cavity appears much fainter because of absorption by the dense, colder
dusty material in the core envelope. However, as one observes at
longer wavelengths (e.g., $\gtrsim 40\:{\rm \mu m}$), the optical
depth is reduced, the far-facing outflow cavity becomes more visible,
and the appearance of the protostar (i.e., the intensity profile along
its outflow axis) becomes more symmetric.

The \textit{Stratospheric Observatory For Infrared Astronomy} (SOFIA)
FORCAST instrument has the ability to observe from MIR wavelengths up
to $\sim$40$\:{\rm \mu m}$ with $\lesssim$3$\arcsec$ angular
resolution. It is thus able to test the above key predictions of Core
Accretion models of massive star formation, i.e., their MIR
morphologies should be aligned with outflow cavities and that at
longer wavelengths the far-facing cavity should become visible as the
overall appearance becomes more symmetric. We note that SOFIA's few
arcsecond resolution at $\sim$40$\:{\rm \mu m}$ means that these
observations are sensitive to FIR morphologies that are induced by the
expected properties of the core infall envelope, rather than by the
disk (also, disks in Competitive Accretion models are expected to be
even smaller than those in Core Accretion models) and that it is the
differences in the predictions of the theoretical formation models on
these core envelope scales that can be tested.

We used SOFIA-FORCAST Early Science observations of the
massive protostar G35.20-0.74 for such a test of the models (Zhang et
al. 2013b). The observations at $37\:{\rm \mu m}$ were able to achieve
a high dynamic range in flux brightness sensitivity of a factor of
$\sim10^4$ and clearly detected the fainter far-facing outflow cavity
at both 31 and 37$\:{\rm \mu m}$. Detailed modeling of the
multiwavelength intensity profiles along the outflow axis, together
with the SED, provided the following constraints on the properties of
the massive protostar, assuming it is the single dominant source of
luminosity: a current stellar mass of $m_* \sim 20 - 34\: M_\odot$,
embedded in a core with $M_c = 240\:M_\odot$, in a clump with
$\Sigma_{\rm cl} \simeq 0.4 - 1\:{\rm g\: cm^{-2}}$.

This work has motivated observations of a larger sample of protostars,
i.e., the SOFIA Massive (SOMA) Star Formation Survey (PI:
Tan). The goal is to observe at least $\sim$50 protostars spanning a
range of environments, evolutionary stages, and core masses. We have
defined four types of sources: Type I: ``MIR sources in IRDCs'' -
relatively isolated sources in Infrared Dark Clouds, some without
detected radio emission; Type II: ``Hyper-compact'' - often jet-like,
radio sources, where the MIR emission extends beyond the observed
radio emission (e.g., G35.20-0.74); Type III: ``Ultra-compact'' -
radio sources where the radio emission is more extended than the MIR
emission; Type IV: ``Clustered sources'' - an MIR source exhibiting
radio emission is surrounded by several other MIR sources within
$\sim$60$\arcsec$. Such classification is somewhat arbitrary, e.g.,
depending on the sensitivity of the MIR or radio continuum
observations, but an evolutionary sequence is expected to hold from
Types I to III. A theoretical calculation of the radio continuum
emission from the early phases of ionization, i.e., of a disk wind
outflow, has been presented by Tanaka, Tan \& Zhang (2016).

Source selection for the SOMA survey mainly utilized the CORNISH
survey of centimeter continuum emission (Hoare et al. 2012),
complemented by radio-quiet MIR sources in IRDCs studied by Butler \&
Tan (2012) and protostars studied at $24\:{\rm \mu m}$ by de Wit et
al. (2009). We included some non-Galactic plane sources and attempted,
where possible, to have a relatively spread-out distribution on the
sky, which aids in the scheduling of SOFIA observations.

In this first paper of the SOMA survey, we present the results of the
first eight sources (including G35.20-0.74), which were observed up to
the end of 2014. These are all Type II sources. Our goal here is to
present the survey data, including public release of the calibrated
images, of these eight sources. We will use these sources to further
test the hypothesis that the appearance of the MIR morphologies of
massive young stellar objects (MYSOs) may be influenced by
outflows. We will also measure the SEDs of the sources and derive
fitting solutions from radiative transfer models, especially the
Zhang, Tan et al. series (hereafter the ZT models) that were
specifically developed for massive protostars. We will also compare
the results of fitting with the more general, commonly used Robitaille
et al. (2007) radiative transfer models. Future papers will carry out
more detailed analyses of images, including outflow axis intensity
profiles, as well as presenting data for additional sources.

\section{Observations}

\subsection{SOFIA data}

The following eight sources, AFGL 4029, AFGL 437, IRAS 07299-1651,
G35.20-0.74, G45.45+0.05, IRAS 20126+4104, Cepheus A and NGC 7538
IRS9, were observed by SOFIA\footnote{SOFIA is jointly operated by the
  Universities Space Research Association, Inc. (USRA), under NASA
  contract NAS2-97001, and the Deutsches SOFIA Institut (DSI) under
  DLR contract 50 OK 0901 to the University of Stuttgart.} (Young et
al. 2012) with the FORCAST instrument (Herter et al. 2013; see
Table~\ref{tab:sofia_obs}). Data were taken on multiple flights
spanning the Early Science period, Cycle 1, and Cycle 2 SOFIA
observing cycles, though typically a single target was observed to
completion on a single flight. All observations were taken at an
altitude between 39,000 and 43,000 ft, which typically yields
precipitable water vapor overburdens of less than 25\,$\mu$m.

FORCAST is a facility imager and spectrograph that employs a Si:As
256$\times$256 blocked-impurity band (BIB) detector array to cover a
wavelength range of 5 to 25\,$\mu$m and a Si:Sb 256$\times$256 BIB
array to cover the range from 25 to 40\,$\mu$m. FORCAST has a dichroic
that allows simultaneous imaging with both arrays, if desired. In
imaging mode the arrays cover a 3.4$\arcmin$$\times$3.2$\arcmin$
instantaneous field of view with 0.768$\arcsec$$^2$ pixels (after
distortion correction). All data were taken by employing the standard
chop-nod observing technique used in the thermal infrared, with chop
and nod throws sufficiently large to sample clear off-source sky.

G35.20-0.74 was observed in the Early Science phase of SOFIA
and was imaged in three filters: 19\,$\mu$m ($\lambda_{\rm
  eff}$=19.7\,$\mu$m; $\Delta\lambda$=5.5\,$\mu$m), 31\,$\mu$m
($\lambda_{\rm eff}$=31.5\,$\mu$m; $\Delta\lambda$=5.7\,$\mu$m), and
37\,$\mu$m ($\lambda_{\rm eff}$=37.1\,$\mu$m;
$\Delta\lambda$=3.3\,$\mu$m). Observations of the rest of the targets
presented here were taken in four filters. For targets observed early
in Cycle 1, namely G45.47+0.05 and IRAS 20126+4104, the SOFIA
11\,$\mu$m ($\lambda_{\rm eff}$=11.1\,$\mu$m;
$\Delta\lambda$=0.95\,$\mu$m) and 25\,$\mu$m ($\lambda_{\rm
  eff}$=25.3\,$\mu$m; $\Delta\lambda$=1.86\,$\mu$m) filters were
employed in the short wavelength camera of FORCAST. After Cycle 1, it
was determined that it would be better to use the 7\,$\mu$m
($\lambda_{\rm eff}$=7.7\,$\mu$m; $\Delta\lambda$=0.47\,$\mu$m)
instead of the 11\,$\mu$m filter because it is closer in wavelength
to the \textit{Spitzer} 8\,$\mu$m filter, which we could use to derive
accurate absolute calibration from the \textit{Spitzer} data. At the
same time we decided to use the 19\,$\mu$m filter instead of the
25\,$\mu$m filter because it is broader and offers better
sensitivity. The two filters used in the long wavelength camera, 31
and 37\,$\mu$m, were used for all Cycle 1 and 2 sources in the
survey. 




SOFIA data were calibrated by the SOFIA pipeline
with a system of stellar calibrators taken across all flights in a
flight series and applied to all targets within that flight series
(see also the FORCAST calibration paper by Herter et
al. 2013). Corrections are made for the airmass of the science targets as
well. The main source of uncertainty in the SOFIA
calibrations is the variability observed in the standard stars'
observed flux throughout the flight and from flight to flight due to
changing atmospheric conditions. The standard deviation of these
measurements will be used as our 1$\sigma$ error on the quoted flux
density measurements, and these are: 2.9\% at 7\,$\mu$m, 1.0\% at
11\,$\mu$m, 3.1\% at 19\,$\mu$m, 5.1\% and 25\,$\mu$m, 3.6\% at
31\,$\mu$m, and 4.6\% at 37\,$\mu$m.

\subsection{\textit{Spitzer} and \textit{Herschel} archival data}

For all objects, data were retrieved from the \textit{Spitzer}
Heritage Archive from all four IRAC (Fazio et al. 2004) channels (3.6,
4.5, 5.8 and 8.0\,$\mu$m). In many cases, the sources in this sample
were so bright that they are saturated in the IRAC images and could
not be used to derive accurate fluxes. Additionally, we incorporated
publicly available imaging observations performed with the
\textit{Herschel Space Observatory}\footnote{\textit{Herschel} is an
  ESA space observatory with science instruments provided by
  European-led Principal Investigator consortia and with important
  participation from NASA. The \textit{Herschel} data used in this
  paper are taken from the Level 2 (flux-calibrated) images provided
  by the \textit{Herschel} Science Center via the NASA/IPAC Infrared
  Science Archive (IRSA), which is operated by the Jet Propulsion
  Laboratory, California Institute of Technology, under contract with
  NASA.} (Pilbratt et al. 2010) and its PACS (Poglitsch et al. 2010)
and SPIRE (Griffin et al. 2010) instruments at 70, 160, 250, 350, and
500\,$\mu$m. The exception is IRAS 07229-6151, for which no
\textit{Herschel} data exist.

\begin{deluxetable*}{ccccccccccc}
\tabletypesize{\scriptsize}
\tablecaption{SOFIA FORCAST Observations: Observation Dates and Exposure Times (s)\label{tab:sofia_obs}}
\tablewidth{0pt}
\tablehead{
\colhead{Source} & \colhead{R.A.(J2000)} & \colhead{Decl.(J2000)} & \colhead{$d$ (kpc)} & \colhead{Obs. Date} & \colhead{7.7$\:{\rm \mu m}$} & \colhead{11.1$\:{\rm \mu m}$} & \colhead{19.7$\:{\rm \mu m}$} & \colhead{25.3$\:{\rm \mu m}$} & \colhead{31.5$\:{\rm \mu m}$} & \colhead{37.1$\:{\rm \mu m}$}
}
\startdata
AFGL 4029 & 03$^h$01$^m$31$\fs$28 & $+$60$\arcdeg$29$\arcmin$12$\farcs$87 & 2.0 & 2014-03-29 & 112 & ... & 158 & ... & 282 & 678 \\
AFGL 437 & 03$^h$07$^m$24$\fs$55 & $+$58$\arcdeg$30$\arcmin$52$\farcs$76 & 2.0 & 2014-06-11 & 217 & ... & 2075 & ... & 2000 & 884 \\
IRAS 07299-1651 & 07$^h$32$^m$09$\fs$74 & $-$16$\arcdeg$58$\arcmin$11$\farcs$28 & 1.68 & 2015-02-06 & 280 & ... & 697 & ... & 449 & 1197 \\
G35.20-0.74 & 18$^h$58$^m$13$\fs$02 & $+$01$\arcdeg$40$\arcmin$36$\farcs$2 & 2.2 & 2011-05-25 & ... & 909 & 959 & ... & 4068 & 4801\\
G45.47+0.05 & 19$^h$14$^m$25$\fs$67 & $+$11$\arcdeg$09$\arcmin$25$\farcs$45 & 8.4 & 2013-06-26 & ... & 309 & ... & 588 & 316 & 585 \\
IRAS 20126+4104 & 20$^h$14$^m$26$\fs$05 & $+$41$\arcdeg$13$\arcmin$32$\farcs$48 & 1.64 & 2013-09-13 & ... & 484 & ... & 1276 & 487 & 1317 \\
Cepheus A & 22$^h$56$^m$17$\fs$98 & $+$62$\arcdeg$01$\arcmin$49$\farcs$39 & 0.70 & 2014-03-25 & 242 & ... & 214 & ... & 214 & 1321 \\
NGC 7538 IRS9 & 23$^h$14$^m$01$\fs$77 & $+$61$\arcdeg$27$\arcmin$19$\farcs$8 & 2.65 & 2014-06-06 & 215 & ... & 653 & ... & 491 & 923
\enddata
\end{deluxetable*}

In addition to using these data for deriving multiwavelength flux
densities of our sources, the \textit{Spitzer} 8\,$\mu$m and
\textit{Herschel} 70\,$\mu$m images are presented for comparison with
our SOFIA images in \S\ref{S:indiv}.

The \textit{Herschel} images, particularly at 70\,$\mu$m, can suffer
from relatively poor image quality due to observations being taken in
fast scanning mode. Point sources are often not circularly symmetric,
and can be severely triangular or square. To enable comparative
morphology as a function of wavelength, the \textit{Herschel}
70\,$\mu$m images were deconvolved to remove most of this asymmetry
and to improve the resolution to be more comparable to the resolution
of SOFIA at 37\,$\mu$m.

\subsection{Data resolutions and deconvolutions}

The resolution of SOFIA through the FORCAST wavelength range
is only slightly dependent upon the effective filter central
wavelength. This is because the image quality is dominated by
in-flight telescope pointing stability, at least at the shorter
wavelengths of FORCAST. The typical resolution achieved for filters
with effective central wavelengths $\lesssim$25\,$\mu$m was about
3$\arcsec$. At wavelengths $\gtrsim$20\,$\mu$m it appears that we are
observing near the diffraction limit. Thus, resolutions presented in
the {\it Spitzer} and SOFIA images in \S\ref{S:indiv} are fairly
similar, i.e., 2.0$\arcsec$ for the \textit{Spitzer} 8\,$\mu$m images,
2.7$\arcsec$ at SOFIA 7\,$\mu$m, 2.9$\arcsec$ at
SOFIA 11\,$\mu$m, 3.3$\arcsec$ at SOFIA 19 and
25\,$\mu$m, 3.4$\arcsec$ at 31\,$\mu$m and 3.5$\arcsec$ at 37\,$\mu$m.

As discussed above, the \textit{Herschel} 70\,$\mu$m images were
deconvolved to improve image quality and resolution. Deconvolution
techniques employ an iterative approach, where the greater the number
of iterations, the better the effective resolution. However, iterating
too much can create artifacts and false structure in the final
deconvolved images. We employed a maximum likelihood approach, using
the max\_likelihood.pro script written by F. Varosi and available in the public IDL
astronomy program database (http:$//$idlastro.gsfc.nasa.gov). We
mildly deconvolved the images (employing no more than 30 iterations),
which tends to correct image PSF abnormalities and create images with
effective resolutions a factor of 1.5-2.0 better than the native image
resolution. Proper deconvolutions require an accurate representation
of the image PSF. Therefore, for each source in our survey, the rest
of the \textit{Herschel} image field was scoured for point sources and
a median combination of all these point sources (after normalization)
was created and used in the deconvolution. The resultant images have
resolutions of 5.0-5.2$\arcsec$, which is $\sim$1.6 times better than
the measured 8.1$\arcsec$ native resolution of \textit{Herschel} at
70\,$\mu$m.

\subsection{Astrometry}

SOFIA observations were performed using the simultaneous observations
with the dichroic in such a way that the relative astrometry between
the four SOFIA images has been determined to be better than a FORCAST
pixel ($\sim$0.77$\arcsec$). The absolute astrometry of the SOFIA data
comes from matching the morphology at the shortest SOFIA wavelength
(either 7 or 11\,$\mu$m) with the \textit{Spitzer} 8\,$\mu$m image (or
shorter IRAC wavelength, if saturated at 8\,$\mu$m). The
\textit{Herschel} 70\,$\mu$m data were found to be off in their
absolute astrometry by up to 5$\arcsec$. For all targets in this
survey, we were able to find multiple sources in common between the
70\,$\mu$m \textit{Herschel} image and sources found in the SOFIA or
\textit{Spitzer} field of view that allowed us to correct the
\textit{Herschel} 70\,$\mu$m absolute astrometry, which is then
assumed to have errors of less than 1$\arcsec$.

\subsection{Other ground-based IR data}

Published and unpublished data from other facilities were also
available for a few sources in our survey and were incorporated into
the SEDs and model fitting (see Table \ref{tab:flux}). 
For G35.20-0.74,
11.7\,$\mu$m (\textit{Si-5}) and 18.3\,$\mu$m (\textit{Qa}) data from
the \textit{Gemini Observatory} T-ReCS instrument (De Buizer \& Fisher
2005) were first published in De Buizer (2006). For IRAS 20126+4104,
\textit{Gemini} T-ReCS 12.5\,$\mu$m (\textit{Si-6}) and 18.3\,$\mu$m
data were also previously published in De Buizer (2007). There are
also previously unpublished \textit{Gemini} T-ReCS 11.7\,$\mu$m and
18.3\,$\mu$m data for IRAS 07299-1651 that we present here. For
G45.47+0.05, we have on hand previously unpublished imaging data from
the \textit{NASA/Infrared Telescope Facility} (IRTF) at
\textit{K} and \textit{L} from the NSFCam instrument (Shure et
al. 1994), as well as previously published (De Buizer et al. 2005)
11.7\,$\mu$m (\textit{N4}) and 20.8 \,$\mu$m (\textit{Q3}) data from
the MIR camera MIRLIN (Ressler et al. 1994).

\section{Analysis Methods}

\subsection{Derivation of Spectral Energy Distributions}

We build SEDs of the eight sources from 3.6\,$\mu$m up to 500\,$\mu$m
with photometric data from \textit{Spitzer}, IRTF,
Gemini, SOFIA and \textit{Herschel}. The
uncertainties are mainly systematic, arising from calibration, which is
in general about 10\%. We used PHOTUTILS, a Python package to measure
the flux photometry.

The position of the protostellar source is generally fixed from
published literature results, e.g., radio continuum emission peak that
is located along a known outflow axis (see \S\ref{S:indiv}). Bright
free-free emission can arise from externally ionized dense clumps, so
ideally confirmation of protostellar location should also be obtained
from high-resolution studies of tracers of hot cores (i.e., warm,
dense gas) and outflows. However, as we discuss in \S\ref{S:indiv},
typically we do not consider that there are large uncertainties in the
source location.

Then, circular apertures of radius $R_{\rm ap}$ are chosen to cover
most of the emission. We try two methods: (1) Fixed Aperture
Radius---the radius is set by considering the morphology of the {\it
  Herschel} 70~$\rm \mu m$ image\footnote{For IRAS 07299-1651 we adopt
  the aperture size based on SOFIA 37$\:{\rm \mu m}$ data since
  no {\it Herschel} data are available.} so as to include most of the
source flux while minimizing contamination from neighboring sources
and (2) Variable Aperture Radius---the radii at wavelengths $<70\:{\rm
  \mu m}$ are varied based on the morphology at each wavelength, again
aiming to minimize contamination from neighboring sources.

The emission at the longer {\it Herschel} wavelengths ($\geq160\:{\rm
  \mu m}$) is typically more extended, which is both a real effect of
the presence of a cooler, massive clump surrounding the protostars,
and also a result of the lower resolution of these data. This is the
main motivation for us to then carry out background subtraction of the
fluxes, based on the median flux density in an annular region
extending from 1 to 2 aperture radii.

Summarizing, for wavelengths $\leq70\:{\rm \mu m}$ the aperture radii
are typically at least several times larger than the beam sizes (and
by greater factors for the fixed aperture method that uses the
$70\:{\rm \mu m}$ aperture radii across all bands). At longer
wavelengths, where the fixed aperture radius set at $70\:{\rm \mu m}$
is always used, the aperture diameter in a few sources (AFGL 4029,
G45.47+0.05, IRAS 20126) begins to become similar to the image
resolutions at the longest wavelengths, i.e., toward $500\:{\rm \mu
  m}$. However, as we will see, for the wavelengths defining the peak
of the SEDs, the source apertures are always significantly larger than
the image resolutions.




\subsection{SED Models and Fitting}

\subsubsection{Zhang \& Tan (ZT) Models}

In a series of papers, Zhang \& Tan (2011), Zhang, Tan \& McKee
(2013), ZTH14 and Y. Zhang \& J. C. Tan (2017, in preparation) have
developed models for the evolution of high- and intermediate-mass
protostars based on the Turbulent Core model (MT03). Zhang \& Tan
(2015) extended these models to treat lower-mass protostars. For
massive star formation, the initial conditions are pressurized, dense,
massive cores embedded in high mass surface density ``clump''
environments. The initial conditions are parameterized by the initial
core mass ($\mc$) and the mean mass surface density of the clump
environment ($\scl$). The latter affects the surface pressure on the
core and therefore, together with $\mc$, determines their sizes and
densities. Cores undergo inside-out collapse (Shu 1977; McLaughlin \&
Pudritz 1996, 1997) with the effect of rotation
described with the solution by Ulrich (1976).

Massive disks are expected to form around massive protostars due to
the high accretion rates. We assume that the mass ratio between the
disk and the protostar is a constant $f_d=m_d/m_*=1/3$, considering
the rise in effective viscosity due to disk self-gravity at about this
value of $f_d$ (Kratter et al. 2008). The disk size is calculated from
the rotating collapse of the core (ZTH14), with the
rotational-to-gravitational energy ratio of the initial core $\beta_c$
set to be 0.02, which is a typical value from observations of low- and
high-mass prestellar cores (e.g., Goodman et al. 1993; Li et al. 2012;
Palau et al. 2013). The disk structure is described with an
``$\alpha$-disk'' solution (Shakura \& Sunyaev 1973), with an improved
treatment to include the effects of the outflow and the accretion
infall to the disk (ZTM13).

Half of the accretion energy is released when the accretion flow
reaches the stellar surface, i.e., the boundary layer luminosity
$L_\mathrm{acc}=Gm_*\dot{m}_*/(2r_*)$, but we assume this part of
luminosity is radiated along with the internal stellar luminosity
isotropically as a single blackbody with total
$L_{*,\mathrm{acc}}=L_*+L_\mathrm{acc}$. This choice is made given the
uncertain accretion geometry near the star, e.g., whether accretion
streamlines are affected by the stellar magnetic field or if the
accretion disk extends all the way in to the stellar surface, and also
given the fact that this emission is at UV/optical/NIR wavelengths and
is reprocessed by dust in the inner regions, including in the outflow,
to longer wavelengths. The other half of the accretion energy is
partly radiated from the disk and partly converted to the kinetic
energy of the disk wind.

The density distribution of the disk wind is described by a
semi-analytic solution, which is approximately a Blandford \& Payne
(1982) wind (see Appendix B of ZTM13), and the mass loading rate of
the wind relative to the stellar accretion rate is assumed to be
$f_w=\dot{m}_w/\dot{m}_*=0.1$, which is a typical value for disk winds
(K\"onigl \& Pudritz 2000). Such a disk wind carves out cavities from
the core, which gradually open up as the protostar evolves. The opening
angle of the outflow cavity is estimated following the method of
Matzner \& McKee (2000) by comparing the wind momentum and that needed
to accelerate the core material to its escape speed (ZTH14). The
accretion rate to the protostar is regulated by this outflow feedback.

The evolution of the protostar is solved using the model by Hosokawa
\& Omukai (2009) and Hosokawa et al. (2010) from the calculated
accretion history. A shock boundary condition is used at very early
stages when the accretion is quasi-spherical. However, then, for most
of the evolution, a photospheric boundary condition is used,
appropriate for disk accretion.


In the above modeling, the evolution of the protostar and its
surrounding structures are all calculated self-consistently from the
two initial conditions of the core: $\mc$ and $\scl$. A third
parameter, the protostellar mass, $\ms$, is used to specify a
particular stage on these evolutionary tracks. In our current model
grid (Y. Zhang \& J. C. Tan, in preparation), $\mc$ is sampled at 10,
20, 30, 40, 50, 60, 80, 100, 120, 160, 200, 240, 320, 400,
480~$M_\odot$,
and $\scl$ is sampled at 0.10, 0.32, 1, 3.2~$\gcm$, for a total of 60
evolutionary tracks. Then along each track, $\ms$ is sampled at 
0.5, 1, 2, 4, 8, 12, 16, 24, 32, 48, 64, 96, 128, and 160~$M_\odot$
(but on each track, the sampling is limited by the final achieved
stellar mass, with star formation efficiencies from the core typically
being $\sim$0.5). There are then, in total, 432 physical models
defined by different sets of $(\mc,~\scl,~\ms)$.

Monte Carlo continuum radiation transfer simulations were performed
for these models using the latest version of the HOCHUNK3d code by
Whitney et al. (2003; 2013). The code was updated to include gas
opacities, adiabatic cooling/heating, and advection (ZTM13). Various
dust opacities are used for different regions around the protostar
(see ZT11), following the choices of Whitney et al. (2003). For each
model, 20 inclinations are sampled evenly in cosine space to produce
the SEDs. To compare with the observations, a variable foreground
extinction $\av$ is applied to the model SEDs. Also, the model SEDs
are convolved with the transmission profiles of the various instrument
filters to estimate flux densities in given observational bands.

We then use $\chisq$ minimization to find the best models to fit a
given set of observations. The reduced $\chisq$ is defined as
\begin{eqnarray}
\chisq & = & \frac{1}{N}\left\{\sum_{\fnumod>\fnuobs}\left[\frac{\log_{10}\fnumod-\log_{10}\fnuobs}
{\sigmau(\log_{10}\fnuobs)}\right]^2\right.\nonumber\\
& + & \left. \sum_{\fnumod<\fnuobs}\left[\frac{\log_{10}\fnumod-\log_{10}\fnuobs}
{\sigmal(\log_{10}\fnuobs)}\right]^2\right\}.
\end{eqnarray}
When $\fnuobs$ is an upper limit, we set $\sigmal=\infty$, i.e.,
there is no contribution to $\chisq$ if $\fnumod<\fnufit$.


For each set of $(\mc,~\scl,~\ms)$, we search for a minimum value of
$\chisq$ by varying the inclination $\inc$ and the foreground
extinction $\av$. The foreground extinction $\av$ is constrained
within a range corresponding to $0.1\:\scl$ to $10\:\scl$, i.e., we
assume that the foreground extinction is somewhat related to that
expected from the ambient clump surrounding the core. We then compare
the minimum $\chisq$ of different cases $(\mc,~\scl,~\ms)$ to find the
best models. Note that for our analysis in this paper we set the
distance to be a known value, based on literature
estimates. Therefore, our SED model grid has only five free
parameters: $\mc$, $\scl$, $m_*$, $\inc$, and $\av$. Our intention is
to then explore to what extent the observed SEDs can be explained by
the different evolutionary stages of a relatively limited set of
initial conditions of massive star formation from the Turbulent Core
Model. We will show the results of the best five models for each
source.

\subsubsection{Robitaille et al. Models}

We also fit the SEDs with the models of Robitaille et al. (2007) for
comparison with the results of the ZT models. To do this, we use the
SED fitting Python package {\it
  sedfitter}\footnote{http://sedfitter.readthedocs.io/en/stable/}
developed by Robitaille et al. (2007). Note that in their fitting code
they adjust the value of the data point to the middle of the error
bar. This influence can be significant when the error bar is large and
asymmetric.

We note that the Robitaille et al. models were developed mostly with
the intention of fitting lower-mass protostars that are typically
observed in lower pressure environments and with lower accretion rates
than the massive protostars of the ZT models. There are $\sim$30
output parameters in the Robitaille et al. models. The key parameters
include
stellar mass, stellar radius, stellar temperature, envelope accretion
rate, envelope outer radius, envelope inner radius, envelope
cavity-opening angle, viewing angle, bolometric luminosity, disk mass,
disk outer radius, disk inner radius, disk accretion rate, extinction
inside the model down to the stellar surface, centrifugal radius,
envelope cavity density, and ambient density around the envelope,
among others. We will show the results for some of these
parameters---those directly comparable with the ZT models---for the
best five Robitaille et al. models.

\subsubsection{General SED Fitting Considerations}

We fit the fiducial SEDs (with fixed aperture size and with
background-subtracted) with the ZT models and the Robitaille et
al. (2007) models. The error bars are set to be the larger of either
10\% of the background subtracted flux density or the value of the
estimated background flux density, for background-subtracted fluxes,
given that order unity fluctuations in the surrounding background flux
are often seen. However, we note that for the protostars analyzed
here, which are relatively bright, background fluxes, especially at
shorter wavelengths and through the peak of the SED, are small
relative to the source. Thus, errors associated with background
subtraction are typically not very significant for our analysis. The
fitting procedure involves convolving model SEDs with the filter
response functions for the various telescope bands. Source distances
were adopted from the literature.  For each source, we present the
five best-fitting models.

Note that short wavelength fluxes, i.e., at $\lesssim8\:{\rm \mu m}$,
may be affected by PAH emission and thermal emission from very small
grains that are transiently heated by single photons. Neither of these
effects are included in the ZT radiative transfer models. Therefore,
we treat the data at these wavelengths as upper limit constraints on
the models.

We also note that the SED model fitting performed here assumes that
there is a single dominant source of luminosity, i.e., effects of
multiple sources, including unresolved binaries, are not accounted
for. This is a general limitation and caveat associated with this
method. Depending on the scales at which apertures are defined and at
which multiple sources may be present, secondary sources may already
be identifiable in the analyzed MIR to FIR images. The SOFIA-FORCAST
data used in this paper have angular resolutions of a few arcseconds,
while the {\it Spitzer}~IRAC $8\:{\rm \mu m}$ images have
$\sim2\arcsec$ resolution. Occasionally, we have access to higher
resolution ground-based MIR imaging of the sources. Future follow-up
observations, e.g., with ALMA and VLA, can also help to
assess the presence of multiple sources.

Finally, both sets of models used in this paper assume smoothly
varying or constant accretion rates. The data being analyzed here were
typically collected within a time frame of about 10 years (i.e., the
{\it Spitzer}, {\it Herschel}, and SOFIA observations). There is
evidence that protostars (e.g., Contreras Pe\~{n}a et al. 2017),
including massive protostars (Caratti O Garatti et al. 2016; Hunter et
al. 2017), can exhibit large luminosity fluctuations, probably due to
bursts of enhanced accretion. However, especially for massive
protostars, the event rate or duty cycle of such burst phases is not
well constrained. Other aspects being equal, one expects that the
luminosity fluctuations of massive protostars will be smaller than for
low-mass protostars, since accretion luminosity makes a smaller
fractional contribution to the total luminosity as protostellar mass
increases (e.g., MT03, ZTH14).



\section{Results}\label{S:results}

\begin{sidewaystable*}
\centering
\setlength{\tabcolsep}{0pt}
\renewcommand{\arraystretch}{0.8}
\vspace{3.5in}
\begin{deluxetable*}{cc|ccc|ccc|ccc|ccc|ccc|ccc|ccc|ccc}
\tabletypesize{\scriptsize}
\tablecaption{Integrated Flux Densities\label{tab:flux}}
\tablewidth{0pt}
\tablehead{
\colhead{Facility} &\colhead{$\lambda$} &\colhead{$F_{\lambda,\rm fix}$}\tablenotemark{a} &\colhead{$F_{\lambda,\rm var}$} \tablenotemark{b} & \colhead{$R_{\rm ap}$} \tablenotemark{c} &\colhead{$F_{\lambda,\rm fix}$} &\colhead{$F_{\lambda,\rm var}$} & \colhead{$R_{\rm ap}$}&\colhead{$F_{\lambda,\rm fix}$} &\colhead{$F_{\lambda,\rm var}$} & \colhead{$R_{\rm ap}$}&\colhead{$F_{\lambda,\rm fix}$} &\colhead{$F_{\lambda,\rm var}$} & \colhead{$R_{\rm ap}$}&\colhead{$F_{\lambda,\rm fix}$} &\colhead{$F_{\lambda,\rm var}$} & \colhead{$R_{\rm ap}$}&\colhead{$F_{\lambda,\rm fix}$} &\colhead{$F_{\lambda,\rm var}$} & \colhead{$R_{\rm ap}$}&\colhead{$F_{\lambda,\rm fix}$} &\colhead{$F_{\lambda,\rm var}$} & \colhead{$R_{\rm ap}$}&\colhead{$F_{\lambda,\rm fix}$} &\colhead{$F_{\lambda,\rm var}$} & \colhead{$R_{\rm ap}$}\\
\colhead{} &\colhead{($\mu$m)} &\colhead{(Jy)} &\colhead{(Jy)} & \colhead{($\arcsec$)} &\colhead{(Jy)} &\colhead{(Jy)} & \colhead{($\arcsec$)} &\colhead{(Jy)} &\colhead{(Jy)} & \colhead{($\arcsec$)} &\colhead{(Jy)} &\colhead{(Jy)} & \colhead{($\arcsec$)} &\colhead{(Jy)} &\colhead{(Jy)} & \colhead{($\arcsec$)} &\colhead{(Jy)} &\colhead{(Jy)} & \colhead{($\arcsec$)} &\colhead{(Jy)} &\colhead{(Jy)} & \colhead{($\arcsec$)} &\colhead{(Jy)} &\colhead{(Jy)} & \colhead{($\arcsec$)} 
}
\startdata
 &  & \multicolumn{3}{c}{AFGL 4029} &  \multicolumn{3}{c}{AFGL 437} &  \multicolumn{3}{c}{IRAS 07299} &  \multicolumn{3}{c}{G35.20-0.74}  &  \multicolumn{3}{c}{G45.47+0.05}  &  \multicolumn{3}{c}{IRAS 20126} &  \multicolumn{3}{c}{Cep A}  &  \multicolumn{3}{c}{NGC 7538 IRS9} \\
\hline
\multirow{2}{*}{IRTF/NSFCam} & \multirow{2}{*}{2.1} & \multirow{2}{*}{-} & \multirow{2}{*}{-} & \multirow{2}{*}{-} & \multirow{2}{*}{-} & \multirow{2}{*}{-} & \multirow{2}{*}{-} & \multirow{2}{*}{-} & \multirow{2}{*}{-} & \multirow{2}{*}{-} & \multirow{2}{*}{-} & \multirow{2}{*}{-} & \multirow{2}{*}{-} & 0.02 & 0.02 & \multirow{2}{*}{7.7} & \multirow{2}{*}{-} & \multirow{2}{*}{-} & \multirow{2}{*}{-} & \multirow{2}{*}{-} & \multirow{2}{*}{-} & \multirow{2}{*}{-} & \multirow{2}{*}{-} & \multirow{2}{*}{-} & \multirow{2}{*}{-} \\
&   &   &   &   &   &   &   &   &   &   &   &   &   & (0.08) & (0.03) &   &   &   &   &   &   &   &   &   &   \\
\hline
\multirow{2}{*}{Spitzer/IRAC} & \multirow{2}{*}{3.6} & 2.60 & 2.14 & \multirow{2}{*}{4.8} & 2.26 & 1.36 & \multirow{2}{*}{12.0} & 1.34 & 1.19 & \multirow{2}{*}{6.0} & 0.50 & 0.30 & \multirow{2}{*}{14.0} & \multirow{2}{*}{-} & \multirow{2}{*}{-} & \multirow{2}{*}{-} & 0.68 & 0.24 & \multirow{2}{*}{4.8} & 15.87 & 6.72 & \multirow{2}{*}{15.0} & 2.80 & 2.15 & \multirow{2}{*}{6.0} \\
&   & (2.71) & (2.23) &   & (2.39) & (1.49) &   & (1.42) & (1.28) &   & (0.60) & (0.34) &   &   &   &   & (0.73) & (0.28) &   & (16.51) & (7.09) &   & (2.99) & (2.24) &   \\
\hline
\multirow{2}{*}{IRTF/NSFCam}	& \multirow{2}{*}{3.8} & \multirow{2}{*}{-} & \multirow{2}{*}{-} & \multirow{2}{*}{-} & \multirow{2}{*}{-} & \multirow{2}{*}{-} & \multirow{2}{*}{-} & \multirow{2}{*}{-} & \multirow{2}{*}{-} & \multirow{2}{*}{-} & \multirow{2}{*}{-} & \multirow{2}{*}{-} & \multirow{2}{*}{-} & 0.12 & 0.08 & \multirow{2}{*}{7.7} & \multirow{2}{*}{-} & \multirow{2}{*}{-} & \multirow{2}{*}{-} & \multirow{2}{*}{-} & \multirow{2}{*}{-} & \multirow{2}{*}{-} & \multirow{2}{*}{-} & \multirow{2}{*}{-} & \multirow{2}{*}{-} \\
&   &   &   &   &   &   &   &   &   &   &   &   &   & (0.08) & (0.08) &   &   &   &   &   &   &   &   &   &   \\
\hline
\multirow{2}{*}{Spitzer/IRAC} & \multirow{2}{*}{4.5} & 3.62 & 2.82 & \multirow{2}{*}{3.6} & 2.77 & 1.89 & \multirow{2}{*}{12.0} & 2.41 & 2.18 & \multirow{2}{*}{6.0} & 1.24 & 0.90 & \multirow{2}{*}{14.0} & 0.25 & 0.18 & \multirow{2}{*}{7.7} & \multirow{2}{*}{-} & \multirow{2}{*}{-} & \multirow{2}{*}{-} & 27.90 & 13.89 & \multirow{2}{*}{15.0} & 7.89 & 5.69 & \multirow{2}{*}{6.0} \\
&   & (3.75) & (2.96) &   & (2.87) & (2.00) &   & (2.54) & (2.33) &   & (1.24) & (0.95) &   & (0.35) & (0.22) &   &   &   &   & (28.67) & (14.65) &   & (8.15) & (5.99) &   \\
\hline
\multirow{2}{*}{Spitzer/IRAC} & \multirow{2}{*}{5.8} & 7.13 & 6.54 & \multirow{2}{*}{7.2} & 10.27 & 5.52 & \multirow{2}{*}{12.0} & 2.96 & 2.53 & \multirow{2}{*}{6.0} & 1.84 & 1.43 & \multirow{2}{*}{14.0} & \multirow{2}{*}{-} & \multirow{2}{*}{-} & \multirow{2}{*}{-} & 1.90 & 0.76 & \multirow{2}{*}{4.8} & 27.61 & 13.51 & \multirow{2}{*}{15.0} & 39.55 & 31.41 & \multirow{2}{*}{6.0} \\
&   & (7.63) & (6.86) &   & (11.05) & (6.19) &   & (3.15) & (2.71) &   & (2.64) & (1.65) &   &   &   &   & (2.24) & (0.92) &   & (30.11) & (14.78) &   & (41.17) & (32.14) &   \\
\hline
\multirow{2}{*}{SOFIA/FORCAST}  & \multirow{2}{*}{7.7} & 12.88 & 12.22 & \multirow{2}{*}{7.7} & 29.60 & 17.04 & \multirow{2}{*}{11.5} & 3.48 & 3.48 & \multirow{2}{*}{7.7} & \multirow{2}{*}{-} & \multirow{2}{*}{-} & \multirow{2}{*}{-} & \multirow{2}{*}{-} & \multirow{2}{*}{-} & \multirow{2}{*}{-} & \multirow{2}{*}{-} & \multirow{2}{*}{-} & \multirow{2}{*}{-} & 21.25 & 11.08 & \multirow{2}{*}{15.0} & 64.24 & 59.12 & \multirow{2}{*}{6.1} \\
&   & (12.72) & (12.28) &   & (28.96) & (18.58) &   & (3.31) & (3.31) &   &   &   &   &   &   &   &   &   &   & (19.00) & (12.02) &   & (62.75) & (59.89) &   \\
\hline
\multirow{2}{*}{Spitzer/IRAC} & \multirow{2}{*}{8.0} & 10.34 & 8.86 & \multirow{2}{*}{7.2} & 24.98 & 13.38 & \multirow{2}{*}{12.0} & 2.30 & 2.12 & \multirow{2}{*}{6.0} & 3.22 & 2.85 & \multirow{2}{*}{14.0} & 0.15 & 0.14 & \multirow{2}{*}{7.7} & 1.36 & 0.44 & \multirow{2}{*}{4.5} & 13.80 & 6.52 & \multirow{2}{*}{15.0} & 41.64 & 27.49 & \multirow{2}{*}{6.0} \\
&   & (11.08) & (9.43) &   & (27.03) & (15.01) &   & (2.51) & (2.29) &   & (4.90) & (3.22) &   & (0.13) & (0.14) &   & (2.04) & (0.64) &   & (17.40) & (7.38) &   & (44.25) & (29.06) &   \\
\hline
\multirow{2}{*}{IRTF/OSCIR} & \multirow{2}{*}{10.5} & \multirow{2}{*}{-} & \multirow{2}{*}{-} & \multirow{2}{*}{-} & \multirow{2}{*}{-} & \multirow{2}{*}{-} & \multirow{2}{*}{-} & \multirow{2}{*}{-} & \multirow{2}{*}{-} & \multirow{2}{*}{-} & \multirow{2}{*}{-} & \multirow{2}{*}{-} & \multirow{2}{*}{-} & 2.37 & 0.07 & \multirow{2}{*}{7.7} & \multirow{2}{*}{-} & \multirow{2}{*}{-} & \multirow{2}{*}{-} & \multirow{2}{*}{-} & \multirow{2}{*}{-} & \multirow{2}{*}{-} & \multirow{2}{*}{-} & \multirow{2}{*}{-} & \multirow{2}{*}{-} \\
&   &   &   &   &   &   &   &   &   &   &   &   &   & (0.38) & (0.24) &   &   &   &   &   &   &   &   &   &   \\
\hline
\multirow{2}{*}{SOFIA/FORCAST}  & \multirow{2}{*}{11.1} & \multirow{2}{*}{-} & \multirow{2}{*}{-} & \multirow{2}{*}{-} & \multirow{2}{*}{-} & \multirow{2}{*}{-} & \multirow{2}{*}{-} & \multirow{2}{*}{-} & \multirow{2}{*}{-} & \multirow{2}{*}{-} & \multirow{2}{*}{-} & \multirow{2}{*}{-} & \multirow{2}{*}{-} & 0.36 & 0.02 & \multirow{2}{*}{7.7} & 0.41 & 0.21 & \multirow{2}{*}{3.2} & \multirow{2}{*}{-} & \multirow{2}{*}{-} & \multirow{2}{*}{-} & \multirow{2}{*}{-} & \multirow{2}{*}{-} & \multirow{2}{*}{-} \\
&   &   &   &   &   &   &   &   &   &   &   &   &   & (0.21) & (0.05) &   & (0.42) & (0.26) &   &   &   &   &   &   &   \\
\hline
\multirow{2}{*}{Gemini/T-ReCS} & \multirow{2}{*}{11.7} & \multirow{2}{*}{-} & \multirow{2}{*}{-} & \multirow{2}{*}{-} & \multirow{2}{*}{-} & \multirow{2}{*}{-} & \multirow{2}{*}{-} & 1.56 & 1.62 & \multirow{2}{*}{1.8} & nan & 2.14 & \multirow{2}{*}{5.0} & 0.32 & 0.36 & \multirow{2}{*}{5.0} & \multirow{2}{*}{-} & \multirow{2}{*}{-} & \multirow{2}{*}{-} & \multirow{2}{*}{-} & \multirow{2}{*}{-} & \multirow{2}{*}{-} & \multirow{2}{*}{-} & \multirow{2}{*}{-} & \multirow{2}{*}{-} \\
&   &   &   &   &   &   &   & (1.71) & (1.66) &   & (3.82) & (2.31) &   & (0.14) & (0.36) &   &   &   &   &   &   &   &   &   &   \\
\hline
\multirow{2}{*}{Gemini/T-ReCS} & \multirow{2}{*}{12.5} & \multirow{2}{*}{-} & \multirow{2}{*}{-} & \multirow{2}{*}{-} & \multirow{2}{*}{-} & \multirow{2}{*}{-} & \multirow{2}{*}{-} & \multirow{2}{*}{-} & \multirow{2}{*}{-} & \multirow{2}{*}{-} & \multirow{2}{*}{-} & \multirow{2}{*}{-} & \multirow{2}{*}{-} & \multirow{2}{*}{-} & \multirow{2}{*}{-} & \multirow{2}{*}{-} & 1.87 & 1.67 & \multirow{2}{*}{6.4} & \multirow{2}{*}{-} & \multirow{2}{*}{-} & \multirow{2}{*}{-} & \multirow{2}{*}{-} & \multirow{2}{*}{-} & \multirow{2}{*}{-} \\
&   &   &   &   &   &   &   &   &   &   &   &   &   &   &   &   & (1.87) & (1.69) &   &   &   &   &   &   &   \\
\hline
\multirow{2}{*}{Gemini/T-ReCS} & \multirow{2}{*}{18.3} & \multirow{2}{*}{-} & \multirow{2}{*}{-} & \multirow{2}{*}{-} & \multirow{2}{*}{-} & \multirow{2}{*}{-} & \multirow{2}{*}{-} & \multirow{2}{*}{-} & \multirow{2}{*}{-} & \multirow{2}{*}{-} & nan & 44.96 & \multirow{2}{*}{7.0} & 2.29 & 4.92 & \multirow{2}{*}{5.0} & 23.84 & 24.12 & \multirow{2}{*}{6.4} & \multirow{2}{*}{-} & \multirow{2}{*}{-} & \multirow{2}{*}{-} & \multirow{2}{*}{-} & \multirow{2}{*}{-} & \multirow{2}{*}{-} \\
&   &   &   &   &   &   &   &   &   &   & (63.03) & (48.00) &   & (2.56) & (4.85) &   & (23.84) & (24.12) &   &   &   &   &   &   &   \\
\hline
\multirow{2}{*}{SOFIA/FORCAST}  & \multirow{2}{*}{19.7} & 57.25 & 54.59 & \multirow{2}{*}{7.7} & 271 & 217 & \multirow{2}{*}{11.5} & 73.82 & 73.82 & \multirow{2}{*}{7.7} & 68.18 & 64.87 & \multirow{2}{*}{14.0} & \multirow{2}{*}{-} & \multirow{2}{*}{-} & \multirow{2}{*}{-} & \multirow{2}{*}{-} & \multirow{2}{*}{-} & \multirow{2}{*}{-} & 138 & 179 & \multirow{2}{*}{24.0} & 172 & 152 & \multirow{2}{*}{6.1} \\
&   & (59.43) & (56.22) &   & (269) & (2219) &   & (74.04) & (74.04) &   & (55.91) & (63.46) &   &   &   &   &   &   &   & (132) & (167) &   & (168) & (154) &   \\
\hline
\multirow{2}{*}{IRTF/MIRLIN}  & \multirow{2}{*}{20.8} & \multirow{2}{*}{-} & \multirow{2}{*}{-} & \multirow{2}{*}{-} & \multirow{2}{*}{-} & \multirow{2}{*}{-} & \multirow{2}{*}{-} & \multirow{2}{*}{-} & \multirow{2}{*}{-} & \multirow{2}{*}{-} & \multirow{2}{*}{-} & \multirow{2}{*}{-} & \multirow{2}{*}{-} & 5.14 & 8.78 & \multirow{2}{*}{7.7} & \multirow{2}{*}{-} & \multirow{2}{*}{-} & \multirow{2}{*}{-} & \multirow{2}{*}{-} & \multirow{2}{*}{-} & \multirow{2}{*}{-} & \multirow{2}{*}{-} & \multirow{2}{*}{-} & \multirow{2}{*}{-} \\
&   &   &   &   &   &   &   &   &   &   &   &   &   & (6.57) & (8.79) &   &   &   &   &   &   &   &   &   &   \\
\hline
\multirow{2}{*}{SOFIA/FORCAST}  & \multirow{2}{*}{25.3} & \multirow{2}{*}{-} & \multirow{2}{*}{-} & \multirow{2}{*}{-} & \multirow{2}{*}{-} & \multirow{2}{*}{-} & \multirow{2}{*}{-} & \multirow{2}{*}{-} & \multirow{2}{*}{-} & \multirow{2}{*}{-} & \multirow{2}{*}{-} & \multirow{2}{*}{-} & \multirow{2}{*}{-} & 45.98 & 45.89 & \multirow{2}{*}{7.7} & 188 & 159 & \multirow{2}{*}{6.4} & \multirow{2}{*}{-} & \multirow{2}{*}{-} & \multirow{2}{*}{-} & \multirow{2}{*}{-} & \multirow{2}{*}{-} & \multirow{2}{*}{-} \\
&   &   &   &   &   &   &   &   &   &   &   &   &   & (33.75) & (42.18) &   & (190) & (163) &   &   &   &   &   &   &   \\
\hline
\multirow{2}{*}{SOFIA/FORCAST}  & \multirow{2}{*}{31.5} & 187 & 171 & \multirow{2}{*}{7.7} & 732 & 656 & \multirow{2}{*}{15.4} & 446 & 446 & \multirow{2}{*}{7.7} & 553 & 502 & \multirow{2}{*}{14.0} & 144 & 135 & \multirow{2}{*}{7.7} & 438 & 352 & \multirow{2}{*}{6.4} & 2771 & 2453 & \multirow{2}{*}{24.0} & 616 & 520 & \multirow{2}{*}{7.7} \\
&   & (194) & (178) &   & (726) & (665) &   & (458) & (458) &   & (551) & (512) &   & (138) & (134) &   & (440) & (365) &   & (2726) & (2466) &   & (620) & (534) &   \\
\hline
\multirow{2}{*}{SOFIA/FORCAST}  & \multirow{2}{*}{37.1} & 405 & 352 & \multirow{2}{*}{7.7} & 878 & 769 & \multirow{2}{*}{15.4} & 681 & 681 & \multirow{2}{*}{7.7} & 1193 & 1061 & \multirow{2}{*}{14.0} & 214 & 189 & \multirow{2}{*}{7.7} & 729 & 528 & \multirow{2}{*}{6.4} & 6262 & 5362 & \multirow{2}{*}{24.0} & 843 & 679 & \multirow{2}{*}{7.7} \\
&   & (419) & (371) &   & (878) & (783) &   & (705) & (705) &   & (1120) & (1071) &   & (202) & (189) &   & (739) & (561) &   & (6275) & (5451) &   & (837) & (699) &   \\
\hline
\multirow{2}{*}{Herschel/PACS}	 & \multirow{2}{*}{70.0} & 350 & 350 & \multirow{2}{*}{11.2} & 1132 & 1132 & \multirow{2}{*}{32.0} & \multirow{2}{*}{-} & \multirow{2}{*}{-} & \multirow{2}{*}{-} & 2628 & 2628 & \multirow{2}{*}{32.0} & 938 & 938 & \multirow{2}{*}{14.4} & 1398 & 1398 & \multirow{2}{*}{12.8} & 14637 & 14637 & \multirow{2}{*}{48.0} & 1568 & 1568 & \multirow{2}{*}{25.6} \\
&   & (394) & (394) &   & (1181) & (1181) &   &   &   &   & (2733) & (2733) &   & (1093) & (1093) &   & (1519) & (1519) &   & (15298) & (15298) &   & (1681) & (1681) &   \\
\hline
\multirow{2}{*}{Herschel/PACS}	 & \multirow{2}{*}{160.0} & 180 & 180 & \multirow{2}{*}{11.2} & 677 & 677 & \multirow{2}{*}{32.0} & \multirow{2}{*}{-} & \multirow{2}{*}{-} & \multirow{2}{*}{-} & 2386 & 2386 & \multirow{2}{*}{32.0} & 622 & 622 & \multirow{2}{*}{14.4} & 655 & 655 & \multirow{2}{*}{12.8} & 10877 & 10877 & \multirow{2}{*}{48.0} & 1019 & 1019 & \multirow{2}{*}{25.6} \\
&   & (264) & (264) &   & (825) & (825) &   &   &   &   & (2807) & (2807) &   & (886) & (886) &   & (783) & (783) &   & (12006) & (12006) &   & (1296) & (1296) &   \\
\hline
\multirow{2}{*}{Herschel/SPIRE} & \multirow{2}{*}{250.0} & 41 & 41 & \multirow{2}{*}{11.2} & 243 & 243 & \multirow{2}{*}{32.0} & \multirow{2}{*}{-} & \multirow{2}{*}{-} & \multirow{2}{*}{-} & \multirow{2}{*}{-} & \multirow{2}{*}{-} & \multirow{2}{*}{-} & 245 & 245 & \multirow{2}{*}{14.4} & 143 & 143 & \multirow{2}{*}{12.8} & \multirow{2}{*}{-} & \multirow{2}{*}{-} & \multirow{2}{*}{-} & 344 & 344 & \multirow{2}{*}{25.6} \\
&   & (104) & (104) &   & (342) & (342) &   &   &   &   &   &   &   & (388) & (388) &   & (210) & (210) &   &   &   &   & (525) & (525) &   \\
\hline
\multirow{2}{*}{Herschel/SPIRE} & \multirow{2}{*}{350.0} & 10.17 & 10.17 & \multirow{2}{*}{11.2} & 75 & 75 & \multirow{2}{*}{32.0} & \multirow{2}{*}{-} & \multirow{2}{*}{-} & \multirow{2}{*}{-} & 429 & 429 & \multirow{2}{*}{32.0} & 61 & 61 & \multirow{2}{*}{14.4} & 25.39 & 25.39 & \multirow{2}{*}{12.8} & 1054 & 1054 & \multirow{2}{*}{48.0} & 91 & 91 & \multirow{2}{*}{25.6} \\
&   & (31.72) & (31.72) &   & (120) & (120) &   &   &   &   & (594) & (594) &   & (113) & (113) &   & (51.61) & (51.61) &   & (1292) & (1292) &   & (177) & (177) &   \\
\hline
\multirow{2}{*}{Herschel/SPIRE} & \multirow{2}{*}{500.0} & 1.16 & 1.16 & \multirow{2}{*}{11.2} & 20.02 & 20.02 & \multirow{2}{*}{32.0} & \multirow{2}{*}{-} & \multirow{2}{*}{-} & \multirow{2}{*}{-} & 127 & 127 & \multirow{2}{*}{32.0} & 8.67 & 8.67 & \multirow{2}{*}{14.4} & 2.93 & 2.93 & \multirow{2}{*}{12.8} & 318 & 318 & \multirow{2}{*}{48.0} & 13.62 & 13.62 & \multirow{2}{*}{25.6} \\
&   & (8.16) & (8.16) &   & (36.77) & (36.77) &   &   &   &   & (196) & (196) &   & (27.61) & (27.61) &   & (11.07) & (11.07) &   & (411) & (411) &   & (52.04) & (52.04) &   \\
\enddata
\tablenotetext{a}{Flux density derived with a fixed aperture size of the 70\,$\rm \mu$m data.}
\tablenotetext{b}{Flux density derived with wavelength-dependent variable aperture sizes.}
\tablenotetext{c}{Aperture radius.}
\tablecomments{The value of flux density in the upper row is derived with background subtraction. The value in the bracket in the lower line is flux density derived without background subtraction.}
\end{deluxetable*}
\end{sidewaystable*}

The SOFIA images for each source are shown below in
\S\ref{S:indiv}.  Also, the type of multiwavelength data available for
each source, the flux densities derived, and the aperture sizes adopted
are listed in Table~\ref{tab:flux}. $F_{\lambda,\rm fix}$ is the flux
density derived with a fixed aperture size and $F_{\lambda,\rm var}$
is the flux density derived with a variable aperture size. The value
of flux density listed in the upper row of each source is derived with
background subtraction, while that derived without background
subtraction is listed in brackets in the lower row.

\subsection{Description of Individual Sources}\label{S:indiv}

Here we describe details about each source as well as presenting their
SOFIA and ancillary imaging data.

\subsubsection{AFGL~4029}


\begin{figure*}
\epsscale{1.15}
\plotone{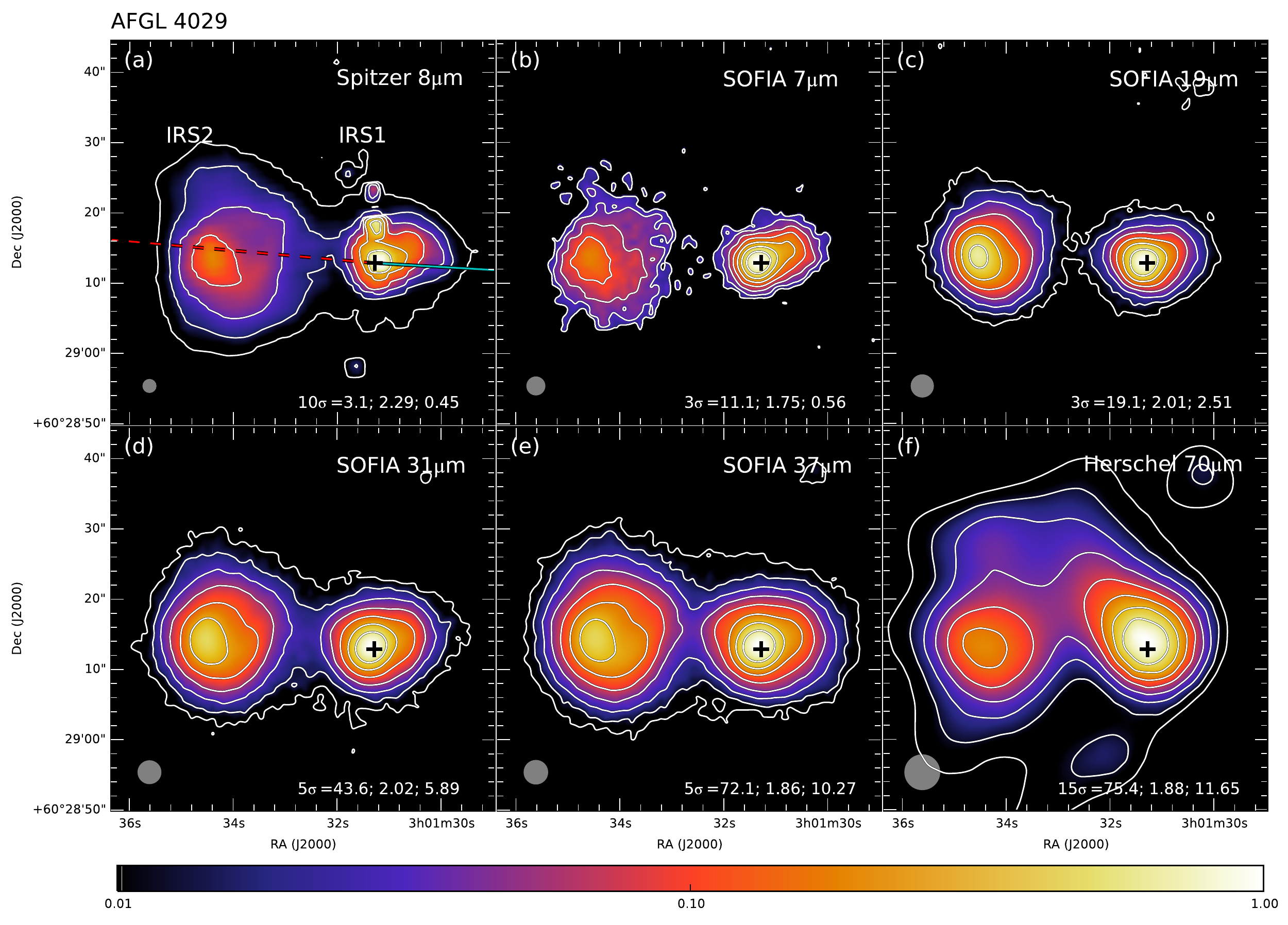}
\caption{
Multiwavelength images of AFGL 4029, 
with facility and wavelength given in the upper right of each panel.
Contour level information is given in the lower right: the lowest
contour level in number of $\sigma$ above the background noise and
corresponding value in mJy per square arcsec, then step size between
each contour in log$_{10}$ mJy per square arcsec, then peak flux in
Jy.  The color map indicates the relative flux intensity compared to
that of the peak flux in each image panel, but only showing the signal
above 3$\sigma$. Gray circles in the lower left show the resolution of
each image. Sources IRS1 (target of interest of this paper) and IRS2
are labeled in panel (a). The black cross in all panels denotes the
position of radio source G138.295+1.555(S) from Zapata et al. (2001)
at R.A.(J2000) = 03$^h$01$^m$31$\fs$28, decl.(J2000) =
$+$60$\arcdeg$29$\arcmin$12$\farcs$87. The lines in panel (a) show the
outflow axis angle, with the solid span tracing the blueshifted
direction and the dotted span tracing the redshifted direction. In
this case, the outflow axis angle is from the H$_2$ and optical jet
emission of Deharveng et al. (1997), and the blueshifted outflow
direction is given by the CO observations of Ginsburg et
al. (2011). In panel (a), the point sources to the north of the
G138.295+1.555(S) position are ghosts in the \textit{Spitzer} image
and should not be interpreted as real
structure.  \label{fig:AFGL4029}}
\end{figure*}

The giant \ion{H}{2} radio region W5 is divided into two subregions,
W5-E and W5-W. W5-E is coincident with the molecular cloud IC 1848A,
and on its eastern border lies the bright infrared region AFGL
4029. Beichman (1979) showed that AFGL 4029 is actually composed of
two mid-IR sources, IRS1 and IRS2, which are separated by
22$\arcsec$. IRS2 appears to be a more evolved \ion{H}{2} region
containing a small stellar cluster dominated by a B1V star (Deharveng
et al. 1997; Zapata et al. 2001). IRS1 is a luminous
($\sim10^4\:L_{\odot}$) and highly reddened ($A_V\sim30$) MYSO
(Deharveng et al. 1997), and has a radio component that has been given
the designation G138.295+1.555 (Kurtz et al. 1994). Later observations
by Zapata et al. (2001) show IRS1 itself to be a binary radio source
with a separation of 0.5$\arcsec$ (or 1000 au given the distance to
the region of about 2.0 kpc (see, e.g.,
Deharveng et al. 2012). Deharveng et al. (1997) detect H$_2$ emission in the NIR
emanating from IRS1 at a position angle of $\sim$265$^{\circ}$, which
is coincident with the high-velocity optical jet seen in [\ion{S}{2}]
(Ray et al. 1990). There also appears to be a smaller
($\sim$1$\arcsec$) radio jet at a similar angle ($\sim$270$^{\circ}$)
to the optical jet (Zapata et al. 2001), as well as a larger, high
energy CO outflow (Ginsburg et al. 2011).

Though IRS1 is the source of interest to this work, both IRS1 and IRS2
are prominently detected in all four wavelengths of SOFIA
(Figure 1). The diffuse and extended nature of IRS2 can be best seen
in the 7\,$\mu$m SOFIA data, consistent with flocculent
morphology seen in the radio continuum maps (Zapata et al. 2001) and
\textit{H} and \textit{K$\arcmin$} images (Deharveng et
al. 1997). IRS1 appears to have a bright peak with a ``tongue'' of
emission extending to the northwest at all SOFIA wavelengths. IRS1 has
been observed at sub-arcsecond resolution in the MIR by Zavagno et
al. (1999, 8--11\,$\mu$m) and de Wit et al. (2009; 24.5\,$\mu$m) and
it appears that this ``tongue'' is an arc-shaped concentration of dust
emission, possibly related to the outflow cavity.


\subsubsection{AFGL~437 (a.k.a. GL~437, G139.909+0.197, IRAS~03035+5819)}

\begin{figure*}
\epsscale{1.15}
\plotone{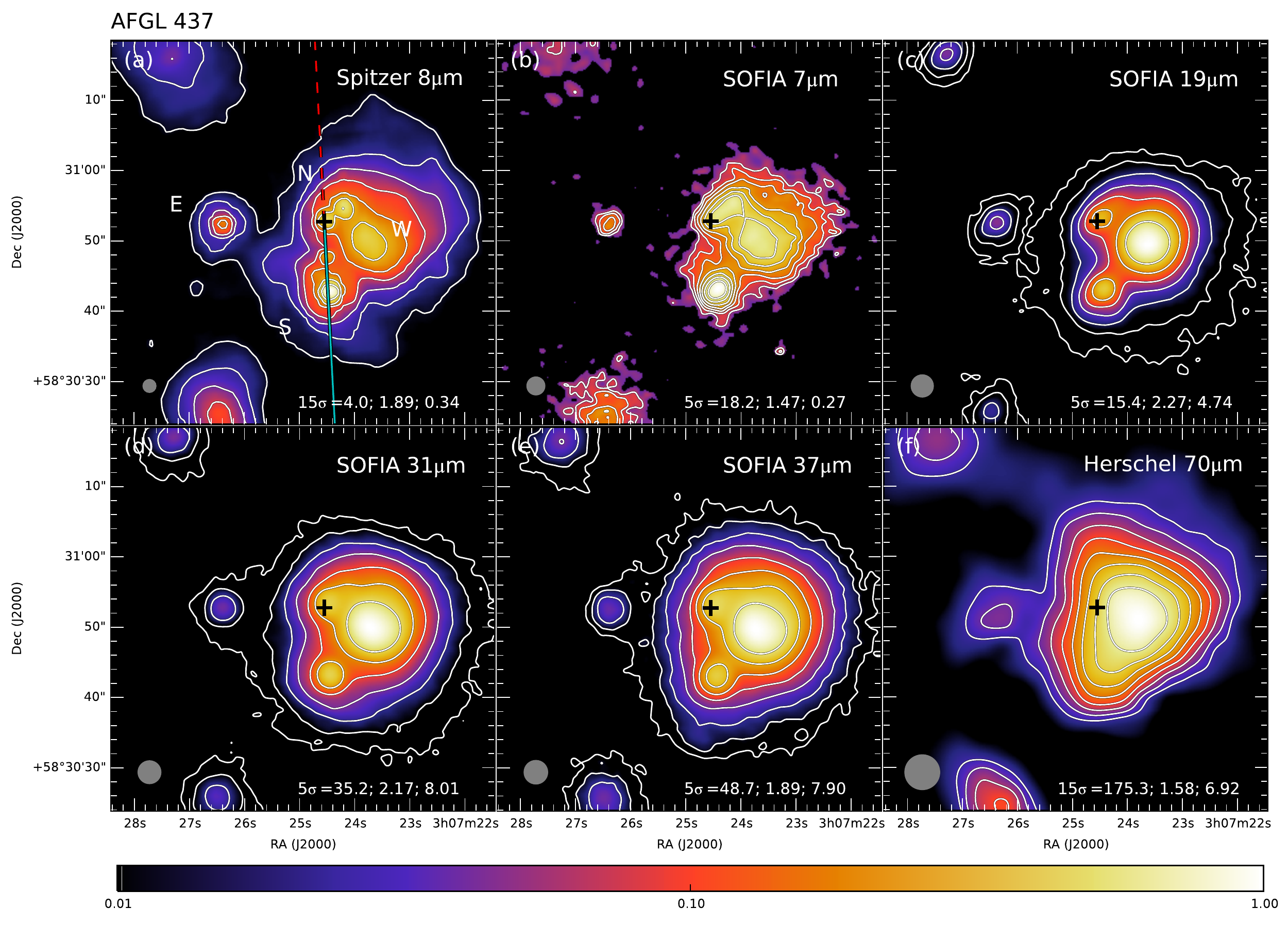}
\caption{
Multiwavelength images of AFGL 437, following the format of
Fig.~\ref{fig:AFGL4029}. The location of the radio continuum source WK34
(Weintraub \& Kastner 1996) is shown as a cross in all panels at
R.A.(J2000) = 03$^h$07$^m$24$\fs$55, decl.(J2000) =
$+$58$\arcdeg$30$\arcmin$52$\farcs$76. The outflow axis angle is from
the NIR bipolar emission angle from Meakin et al. (2005), and the
blueshifted outflow direction is given by the CO observations of
G\'omez et al. (1992).\label{fig:AFGL437}}
\end{figure*}

AFGL~437 is a compact infrared cluster (Wynn-Williams et al. 1981;
Weintraub \& Kastner 1996) that is dominated by four bright sources
named AFGL~437~N, S, E, and W. Based on a combination of kinematic and
spectroscopic distance measurements, Arquilla \& Goldsmith (1984)
estimated the distance to this region to be 2.0~kpc, and the total
luminosity of the cluster is estimated to be
$\sim3\times10^4\:L_{\odot}$. Radio centimeter continuum emission was first
detected from two of the sources, with most of the emission coming
from source W (determined to be an \ion{H}{2} region), with some weak
emission coming from source S (Torrelles et al. 1992). In the
infrared, Weintraub \& Kastner (1996) found that source N could be
resolved into two components, with the southeastern source of the
two, dubbed WK~34, found to be the most embedded source in the
cluster, and also associated with weak radio continuum emission.

This cluster of infrared sources is at the center of a CO molecular
outflow (G\'omez et al. 1992; Qin et al. 2008) that is roughly
oriented north-south and poorly collimated, making it difficult to
accurately determine which source(s) might be driving the
outflow. Weintraub \& Kastner (1996) found the cluster to be
surrounded by an infrared reflection nebula that has a polarization
pattern centro-symmetric with respect to source WK~34, which they
believe traces an outflow cavity from that source. Kumar~Dewangan \&
Anandarao (2010) resolve a finger-shaped ``green fuzzy'' emission
region extending north from WK~34 in the \textit{Spitzer} IRAC images,
which they speculate is tracing H$_2$ emission from an outflow lobe
(though such emission is not a dependable outflow tracer; see De
Buizer \& Vacca 2010 and Lee et al. 2013). Perhaps the most convincing
evidence of an outflow from WK~34 comes from the \textit{Hubble}
NICMOS polarimetric imaging of this source (Meakin et al. 2005), which
resolves a well-collimated bipolar reflection nebula that is oriented
north-south and consistent with the outflow observations described
above. If this is the main source of outflow, previous SED modeling of
WK~34 yields an estimated source mass and luminosity to be
$\sim7\:M_{\odot}$ and $\sim1\times10^3\:L_{\odot}$, respectively
(Kumar~Dewangan \& Anandarao 2010), which is more consistent with an
intermediate-mass object than a true MYSO. We will see below that one
of the favored ZT radiative transfer models includes a source with
$m_*=8\:M_\odot$, although higher-mass cases are allowed.

In the SOFIA data, we barely resolve source AFGL~437~N at
7\,$\mu$m into WK~34 and its companion, but they are resolved in the
\textit{Spitzer} 8\,$\mu$m data (Figure 2). We see no evidence of
infrared emission to the north of WK~34 in the SOFIA data,
which is where the green fuzzy emission has been seen. However, if the
larger-scale CO outflow is being driven by WK~34, observations by
G\'omez et al. (1992) and Qin et al. (2008) show that the blueshifted
outflow lobe should be to the south. The expectation would be that we
should see the blueshifted outflow cavity more readily due to
decreased extinction. Unfortunately, any southern outflow cavity from
WK~34 cannot be discerned from the SOFIA data due to the
resolution of the observations and the close proximity of source S to
the south. However, the sub-arcsecond resolution 24.5\,$\mu$m images
from de Wit et al. (2009) conclusively show that there is no extended
emission south of WK~34 at that wavelength (at least to within their
detection limit).

Interestingly, the source with the peak infrared brightness is
AFGL~437~S at the shorter MIR wavelengths, but at wavelengths longer
than 19\,$\mu$m, the UC \ion{H}{2} region AFGL~437~W is where the
brightness peaks (see also de Wit et al. 2009), perhaps further
indicating that WK~34 is not an MYSO.

\subsubsection{IRAS~07299-1651 (a.k.a. AFGL~5234, S302, DG~121, RCW~7, G232.62+01.00)}

\begin{figure*}
\epsscale{1.15}
\plotone{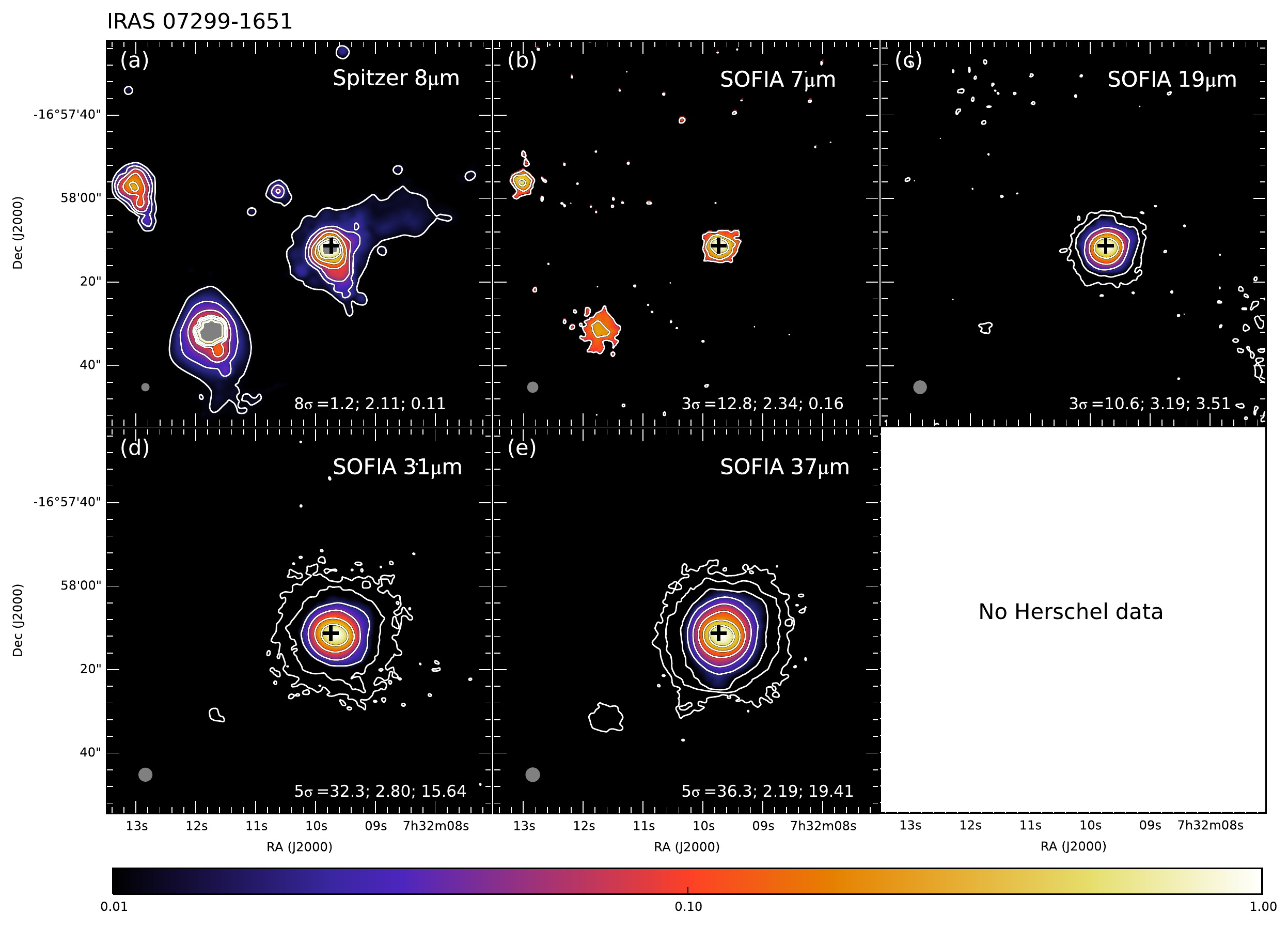}
\caption{
Multiwavelength images of IRAS 07299-1651, following the format of
Fig.~\ref{fig:AFGL4029}. The gray areas in panel (a) are where the
sources have saturated in the IRAC image. Also in panel (a) there are
extensions to the southwest of the three brightest sources, which are
ghosts that should not be interpreted as real structure. The location
of the radio continuum source of Walsh et al. (1998) is shown as a
cross in all panels at R.A.(J2000) = 07$^h$32$^m$09$\fs$74,
decl.(J2000) = $-$16$\arcdeg$58$\arcmin$11$\farcs$28. There are no
outflow maps from which to discern an outflow angle or direction for
this source. \label{fig:IRAS07299}}
\end{figure*}


\begin{figure*}
\epsscale{1.00}
\plotone{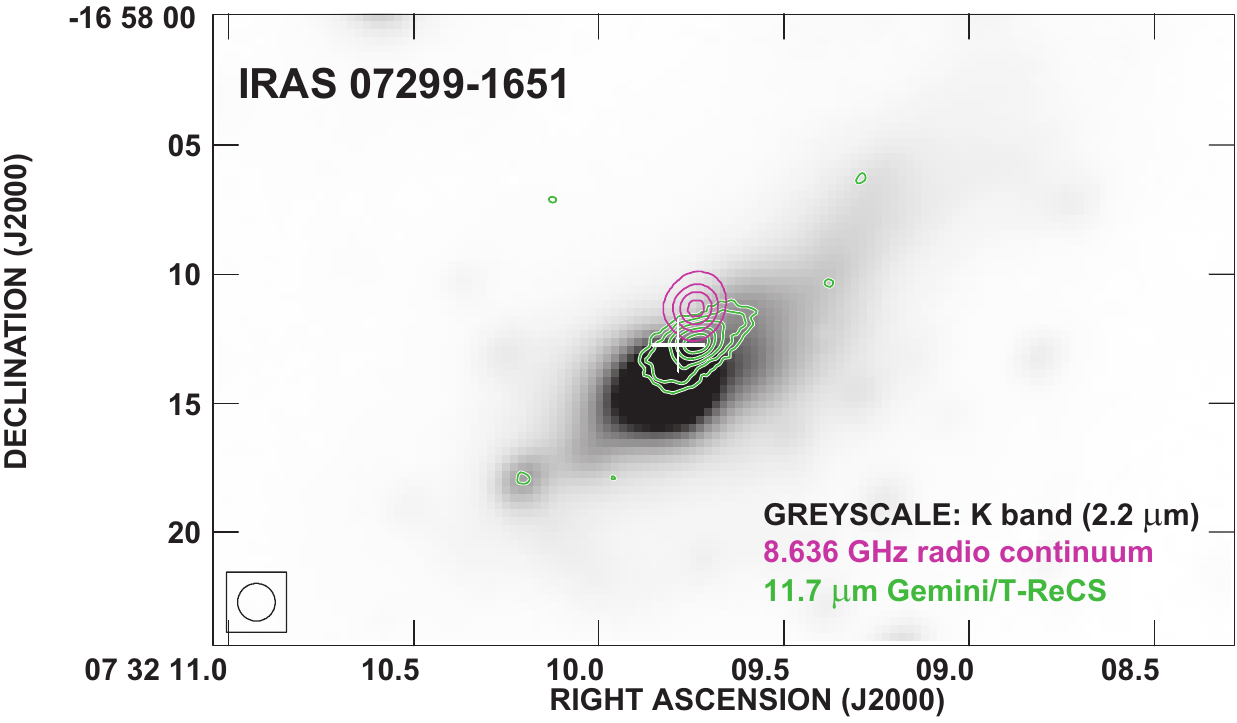}
\caption{
Image of IRAS 07299-1651 comparing the 11.7um Gemini/T-ReCS image
(green contours) with the near-infrared (grayscale) and radio
continuum (red contours) emission, as well as the methanol maser location
(white cross) from Walsh et al. (1999). \label{fig:IRAS07299extra}}
\end{figure*}

Figure~\ref{fig:IRAS07299} presents our standard multiwavelength data
for IRAS~07299-1651. The NIR emission from this source was shown to
have a compact center with diffuse emission extended at a position
angle of 305$^{\circ}$ (Walsh et al. 1999). Follow-up observations in
the MIR in the $N$-band ($\sim$10\,$\mu$m) by Walsh et al. (2001) with
the ESO Max Planck Institute 2.2~m telescope show a compact, perhaps
slightly elongated source at this location. Our Gemini South
8~m observations at 11.7\,$\mu$m at higher resolution and sensitivity
show an elongated appearance resembling the NIR morphology, with a
compact core and extended diffuse emission (see
Figure~\ref{fig:IRAS07299extra}). However, the MIR emission is not
coincident with the NIR emission, and neither is coincident with the
radio continuum peak of Walsh et al. (1998). The peak in emission in
the \textit{Spitzer} 8\,$\mu$m image (Figure~\ref{fig:IRAS07299}a) is
coincident with the peak in the 11.7\,$\mu$m Gemini image to
within the accuracies of our astrometry
($\lesssim$0.5$\arcsec$)\footnote{This is different from the location
  of the peak seen in the $N$-band image of Walsh et al. (2001), which is
  likely in error.}. As one looks to shorter wavelengths in the
\textit{Spitzer} IRAC data, the peak moves closer and closer to the
2\,$\mu$m peak location, suggesting that extinction might be playing a
role. At the resolution of SOFIA, the object looks rather point like,
with a possible extension of the emission to the northwest seen at 31 and
37\,$\mu$m (Figure~\ref{fig:IRAS07299}(d) and (e)).

Given the extended nature of the NIR and MIR emission of this target
at high angular resolution, it was deemed a good candidate for being
morphologically influenced by an outflow. The hypothesis is that the
radio continuum source also drives an outflow, and the extended NIR
and MIR emission are coming from the blueshifted outflow cavity. To
date, however, there are no maps of outflow indicators of this source
from which we may derive an outflow axis. Evidence of an outflow from
this region does exist, including spectra that show that the $^{12}$CO
gas is considered to be in a ``high-velocity" state (Shepherd \&
Churchwell 1996). Liu et al. (2010) mapped the integrated $^{13}$CO
emission at $\sim$1$\arcmin$ resolution and found it to be extended
parallel and perpendicular to the NIR/MIR extension on the scale of
$\sim$4$\arcmin$ in each direction. No velocity maps are presented in
that work, and they claim that the emission is tracing a molecular
core (not outflow), from which they estimate a gas mass of
1.2$\times$10$^3\:M_{\odot}$.

De Buizer (2003) claimed that in some cases the groupings of 6.7 GHz
methanol maser spots may lie in an elongated distribution that is
parallel to the outflow axis for some MYSOs. Fujisawa et al. (2014)
showed that the 6.7 GHz methanol maser spots are distributed over two
groupings separated by about 60 mas with a total distributed area of
about 20 mas $\times$ 70 mas (or 40 au $\times$ 120 au, given the
distance of 1.68~kpc estimated from the trigonometric parallax
measurements of the 12~GHz methanol masers present in this source by
Reid et al. 2009). Though there are two groups of masers, they have a
velocity gradients along their shared axis of elongation and are
distributed at a position angle of 340$^{\circ}$.

\subsubsection{G35.20-0.74 (a.k.a. IRAS~18566+0136)}\label{S:G35.2}

\begin{figure*}
\epsscale{1.15}
\plotone{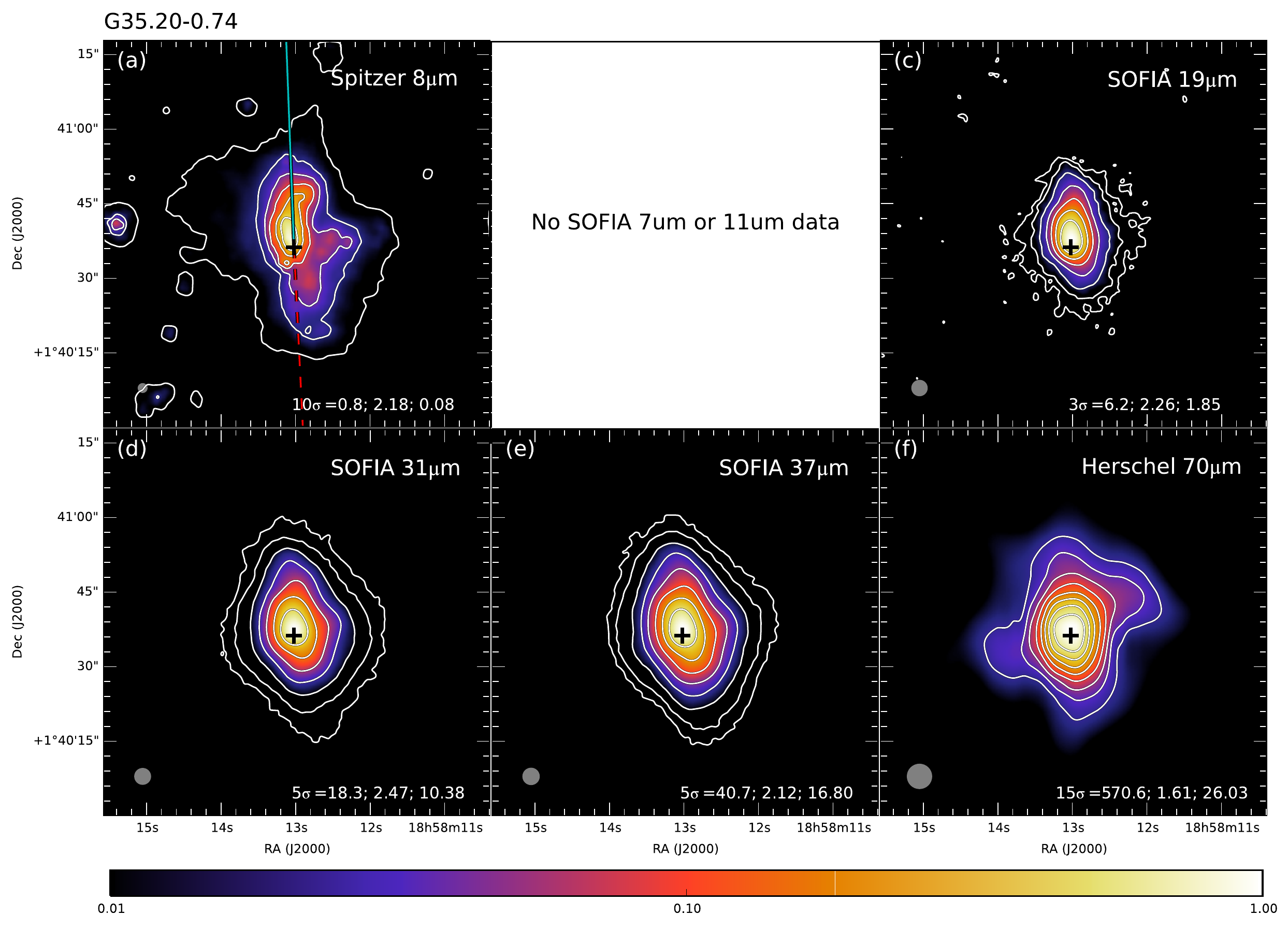}
\caption{
Multiwavelength images of G35.20-0.74, following the format of
Fig.~\ref{fig:AFGL4029}. The location of radio continuum source 7 from
Gibb et al. (2003) is shown as a cross in all panels at R.A.(J2000) =
18$^h$58$^m$13$\fs$02, decl.(J2000) =
$+$01$\arcdeg$40$\arcmin$36$\farcs$2. In panel (a) the axis of the
radio jet is shown (Gibb et al. 2003); the blueshifted 
direction is derived from CO observations of Birks et
al. (2006).\label{fig:G35.20-0.74}}
\end{figure*}

The G35.20-0.74 star forming region, at a distance of 2.2~kpc (Zhang
et al. 2009; Wu et al. 2014), was first identified as a star-forming
molecular cloud through ammonia observations by Brown et
al. (1982). Dent et al. (1985a) were the first to resolve the emission
in this region into a molecular ridge running northwest to southeast
seen in CS(2-1), with a nearly perpendicular outflow seen in
CO(1-0). Dent et al. (1985b) found the NIR emission to be coming from
an elongated north-south distribution. Heaton \& Little (1988)
observed this region in centimeter radio continuum and were able to resolve
three compact sources arranged north-south, and concluded that the
central source was likely a UC~\ion{H}{2} region while the north and
south sources had spectral indices consistent with free-free emission
from a collimated, ionized, bipolar jet. The orientation of this jet
(P.A.$\sim$2$^{\circ}$) appears to be different from that of the CO
outflow (P.A.$\sim$58$^{\circ}$), which has been interpreted either as
evidence for precession of the ionized jet (Heaton \& Little 1988;
Little et al. 1998; S{\'a}nchez-Monge et al. 2014; Beltr\'an et
al. 2016), or multiple outflows from multiple sources (Gibb et
al. 2003; Birks et al. 2006).

G35.20-0.74 was the first source observed among those in the SOMA
survey sample, and the SOFIA-FORCAST imaging data were presented
by Zhang et al. (2013b). These data helped define the infrared SED of
the source, which implied an isotropic luminosity of
$3.3\times10^4\:L_\odot$. However, modeling the emission (with early
versions of the ZT radiative transfer models that had fixed outflow
cavity-opening angles; ZTM13), including 10 to 40$\:{\rm \mu m}$
intensity profiles, as being due to a single protostar driving an
outflow along the N-S axis, Zhang et al. (2013b) derived a true
bolometric luminosity in the range $\sim(0.7-2.2)\times10^5\:L_\odot$,
i.e., after correcting for foreground extinction and anisotropic
beaming. Note, these estimates were based on a limited, ad hoc
exploration of model parameter space. They correspond to protostellar
masses in the range $m_*\simeq$ 20 to 34$\:M_\odot$ accreting at rates
$\dot{m}_*\sim10^{-4}\:M_\odot\:{\rm yr}^{-1}$ from cores with initial
mass $M_c=240\:M_\odot$ in clump environments with $\Sigma_{\rm
  cl}=0.4$ to $1.0\:{\rm g\:cm^{-2}}$ and with foreground extinctions
from $A_V=0$ to 15~mag.

Such an interpretation of outflow orientation is broadly consistent
with the sub-arcsecond VLA observations of this field by Gibb et
al. (2003) at centimeter wavelengths, which show that the three
concentrations of radio continuum emission from Heaton \& Little
(1988) break up into 11 individual knots all lying along a
north-south position angle. The central source itself is resolved into
two sources separated by 0.8$\arcsec$. The northern of the two central
sources (source 7) has a spectral index typical of a UC~\ion{H}{2}
region and was claimed by Gibb et al. to be the most likely driving
source of the radio jet. Beltr\'an et al. (2016) have also identified
this source, a component of a binary system they refer to as 8a, as
the likely driving source.  To be able to ionize the UC~\ion{H}{2}
region, Beltr\'an et al. (2016) estimate that it have the H-ionizing
luminosity of at least that of a spectral type B1 zero age main
sequence (ZAMS) star. This radio source is coincident with Core B of
S{\'a}nchez-Monge et al. (2014) seen at 870\,$\mu$m with ALMA (which
is the same as source MM1b from the 880\,$\mu$m SMA observations of
Qiu et al. 2013), who estimate the core mass in this vicinity to be
18$\:M_{\odot}$.





The scenario of north-south directed protostellar outflows is also
supported by MIR imaging. High-resolution MIR images of this region by
De Buizer (2006) showed that the emission is peaked to the north of
radio source 7 and elongated in a north-south orientation, very
similar to what was seen in the NIR for the first time by Dent et
al. (1985b). A weak extended area of emission was seen to the south,
and can be seen in the much more sensitive \textit{Spitzer} 8\,$\mu$m
data (Figure~\ref{fig:G35.20-0.74}a). The outflow/jet is blueshifted
to the north (e.g., Gibb et al. 2003; Wu et al. 2005) and is likely to
be the reason why we see emission predominantly from that side of
source 7 at shorter MIR wavelengths. However, as discussed by Zhang et
al. (2013b), the longer wavelength SOFIA images (Figure
\ref{fig:G35.20-0.74}) are able to detect emission also from the
southern, far-facing outflow cavity.



Finally, we note that for G35.20-0.74 we could not derive an accurate
background-subtracted flux density for the Gemini data with the fixed
aperture size due to the small size of the images. Thus, in this case
we estimate a background-subtracted flux density derived from a
smaller aperture size.

\subsubsection{G45.47+0.05}

\begin{figure*}
\epsscale{1.15}
\plotone{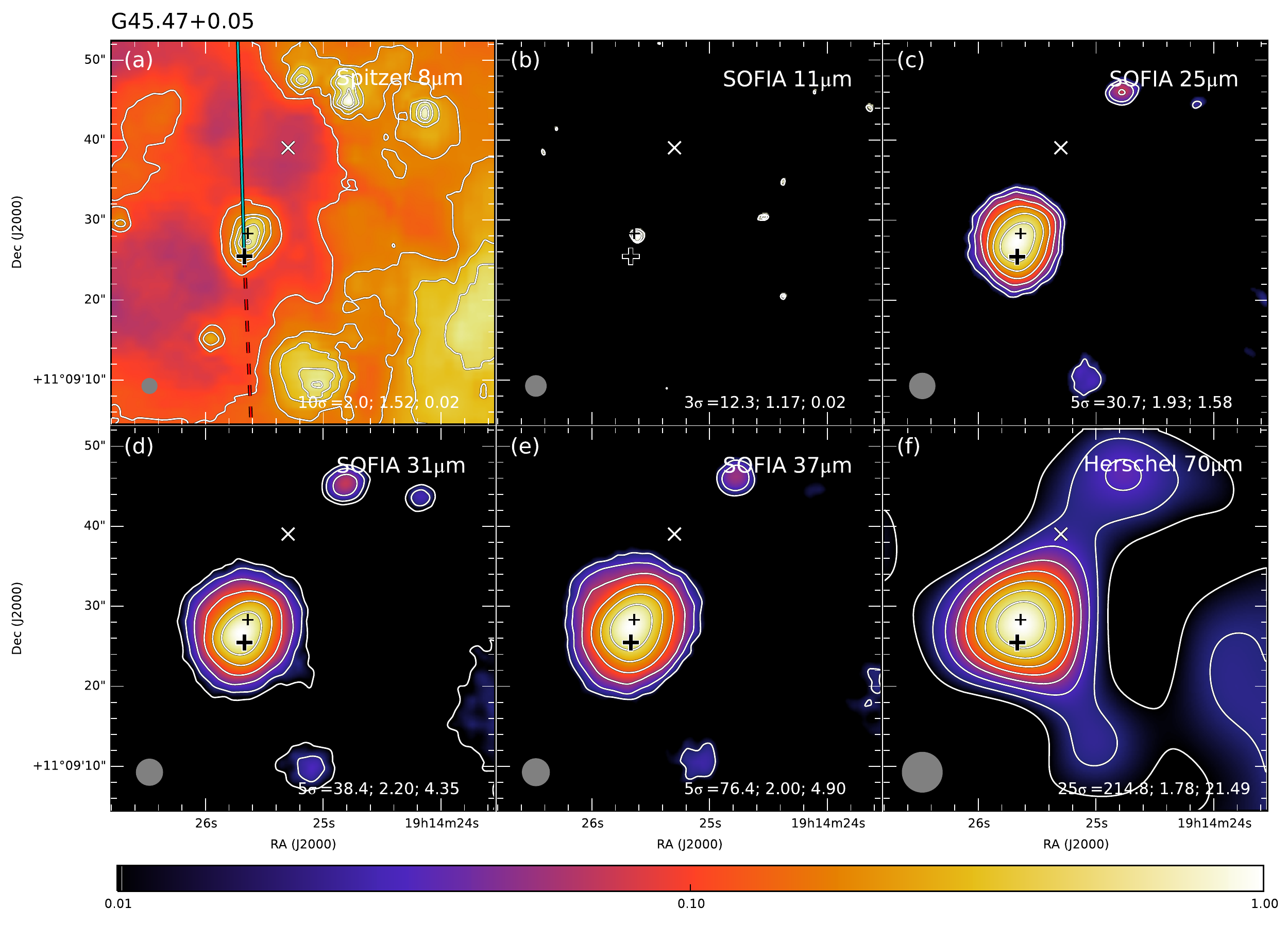}
\caption{
Multiwavelength images of G45.47+0.05, following the format of
Fig.~\ref{fig:AFGL4029}. The location of the 6~cm radio continuum peak
of the UC~\ion{H}{2} region of White et al. (2005) is shown as a large
cross in all panels at R.A.(J2000) = 19$^h$14$^m$25$\fs$67,
decl.(J2000) = $+$11$\arcdeg$09$\arcmin$25$\farcs$45. The location of
the 2MASS source J19142564+1109283 is shown by the small cross. The
location of the peak of the blueshifted SiO(2-1) emission of Wilner et
al. (1996) is shown by an $\times$. The outflow axis angle and the
blueshifted outflow direction are given by the HCO$^+$ observations
of Wilner et al. (1996).
\label{fig:G45.47+0.05}}
\end{figure*}

\begin{figure*}
\epsscale{1.00}
\plotone{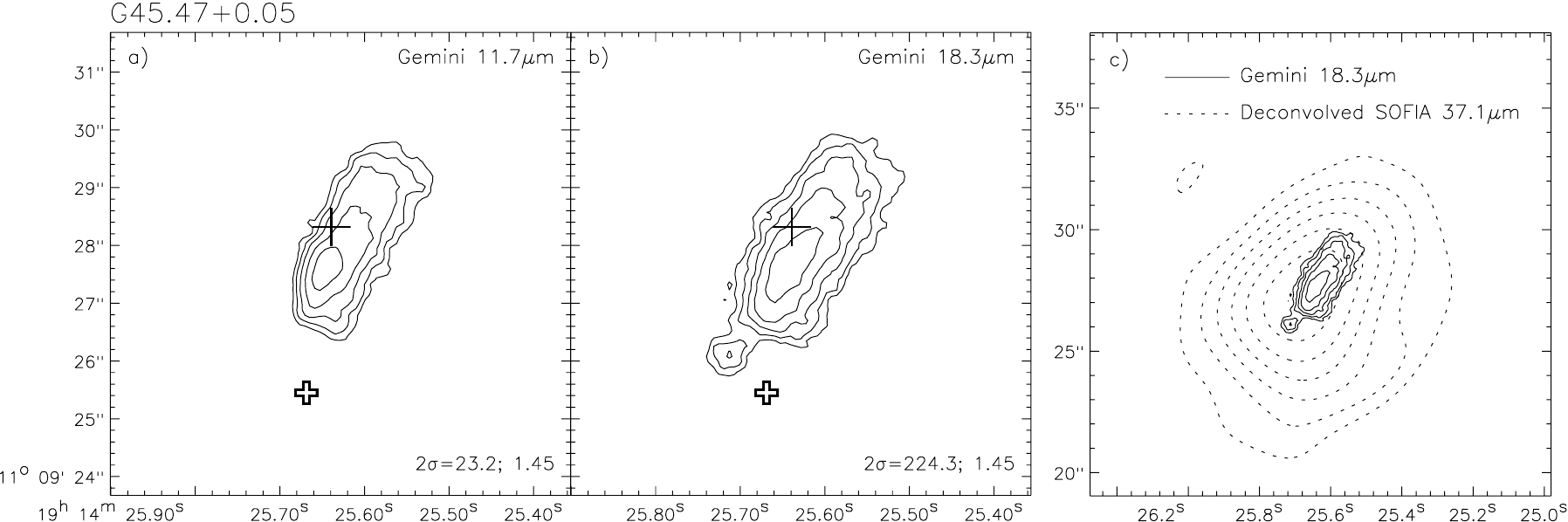}
\caption{
Sub-arcsecond resolution MIR images of G45.47+0.05 from
Gemini T-ReCS. Symbols and annotation are the same as in
Figure~\ref{fig:G45.47+0.05}.\label{fig:G45.47+0.05extra}}
\end{figure*}

G45.47+0.05 was first detected as a UC~\ion{H}{2} region in the radio
continuum at 6~cm (Wood \& Churchwell 1989) and lies at a distance of
8.4~kpc, based upon the trigonometric parallax measurements of masers
in nearby G45.45+0.05 (Wu et al. 2014). G45.47+0.05 has a relatively
high luminosity ($\sim10^6\:L_{\odot}$; Hern\'andez-Hern\'andez et
al. 2014), testifying to its nature as an MYSO. The UC~\ion{H}{2} region
is also coincident with other MYSO tracers like hydroxyl and water
masers (Forster and Caswell 1989).

There is some debate as to the nature of the outflow and driving
source in this region. \textit{Spitzer} IRAC images show a source that
is a bright ``green fuzzy,'' and consequently was categorized as being
a ``likely MYSO outflow candidate'' in the work of Cyganowski et
al. (2008). However, Lee et al. (2013) find no H$_2$ emission
component to the green fuzzy and classify the NIR emission as a
reflection nebula (possibly from an outflow cavity). This region was
mapped in HCO$^+$(1-0), a potential outflow indicator, by Wilner et
al. (1996), who showed that the emission is oriented roughly
north-south (P.A.$\sim$3$^{\circ}$) and centered on the location of
the UC~\ion{H}{2} region, with blueshifted emission to the
north. They also mapped the area in another outflow indicator,
SiO(2-1), and find emission at the location of the UC~\ion{H}{2}
region with a single blueshifted component lying $\sim$14$\arcsec$ to
the northwest at a position angle of about -25$^{\circ}$ (see Figure
\ref{fig:G45.47+0.05}). However, Ortega et al. (2012) mapped the area
in $^{12}$CO(3-2) and found the red and blueshifted peaks to be
oriented at an angle of $\sim$15$^{\circ}$, but with an axis offset
$\sim$10$\arcsec$ southeast of the UC~\ion{H}{2} region.

The observations of De Buizer et al. (2005) first showed that the MIR
emission in this region is offset $\sim$2.5$\arcsec$ northwest of the
radio continuum peak. \textit{Spitzer} IRAC and 2MASS data confirm
this offset of the peak of the NIR/MIR emission, and show a similar
extended morphology, with the axis of elongation oriented at a
position angle of about -30$^{\circ}$ and pointing radially away from
the radio continuum peak. The SOFIA data (Figure
\ref{fig:G45.47+0.05}) show this same morphology at wavelengths
greater than 19\,$\mu$m (the 11\,$\mu$m SOFIA observation is a shallow
integration that only barely detects the peak emission from the
source). We also present higher angular resolution Gemini T-ReCS
imaging at 11.7 and 18.3~$\rm \mu m$ in
Figure~\ref{fig:G45.47+0.05extra}, which also shows this offset and
elongation. We note that the elongated morphology persists out to even
longer wavelengths, as seen in both the \textit{Herschel} 70\,$\mu$m
data, as well as {\it JCMT} SCUBA images at 850\,$\mu$m
(Hern\'andez-Hern\'andez et al. 2014).

There are two main scenarios to describe the outflow and driving
source in this region. The first is that the massive star(s) powering
the UC~\ion{H}{2} region is (are) also driving a roughly north-south
outflow, with the CO, HCO$^+$, and SiO emission tracing different
parts of the wide-angled outflow. The NIR and MIR emission are
emerging from the blueshifted outflow cavity. The slight offset
between the UC~\ion{H}{2} region peak and the NIR/MIR emission may be
due to the high extinction toward the UC~\ion{H}{2} region
itself. High spatial resolution adaptive optics imaging in the NIR of
this source (Paron et al. 2013) show it to be a triangular-shaped
emission region, with its southern apex pointing directly back at the
UC~\ion{H}{2} region location. The opening angle of this outflow cone
is $\sim$50$^{\circ}$, with its axis of symmetry pointing toward the
blueshifted SiO emission, hinting that this might be a cone-shaped
outflow cavity/reflection nebula emanating from the UC~\ion{H}{2}
region. Furthermore, while the SOFIA 11\,$\mu$m emission is peaked
close to the MIR and NIR peaks seen by \textit{Spitzer} and 2MASS, the
peak of the longer wavelength MIR emission is centered closer to the
UC~\ion{H}{2} region peak, as would be expected in this scenario. It
is not clear that we are detecting any additional emission from the
redshifted outflow cavity, even at the longest SOFIA wavelengths.

The second scenario is that the outflow is coming from an NIR star at
the western apex of the triangular-shaped NIR-emitting region seen in
the adaptive optics images of Paron et al. (2013). They dub this
source 2MASS~J19142564+1109283 (see Figure \ref{fig:G45.47+0.05}a),
which is actually the name of the entire NIR emitting region
(2MASS did not have the resolution to separate this stellar
source from the rest of the extended emission). In this scenario, the
outflow cone from 2MASS~J19142564+1109283 would have a much wider
opening angle of about $\sim$90$^{\circ}$ and have an axis of symmetry
that points toward the blueshifted $^{12}$CO(3-2) peak seen by
Ortega et al. (2012). This scenario is not favored here because it
does not explain the location of the southern redshifted $^{12}$CO
outflow peak, which would be at an angle $\sim$80$^{\circ}$ from the
outflow axis, nor does it explain the roughly north-south outflow
emission seen in HCO$^+$(1-0) and SiO(2-1).

Whether the driving source is a stellar object at the center of the
UC~\ion{H}{2} region or 2MASS~J19142564+1109283, it appears that the
MIR emission observed in the region is coming from a blueshifted
outflow cavity.

\subsubsection{IRAS~20126+4104 (a.k.a. G078.12+03.64)}

\begin{figure*}
\epsscale{1.15}
\plotone{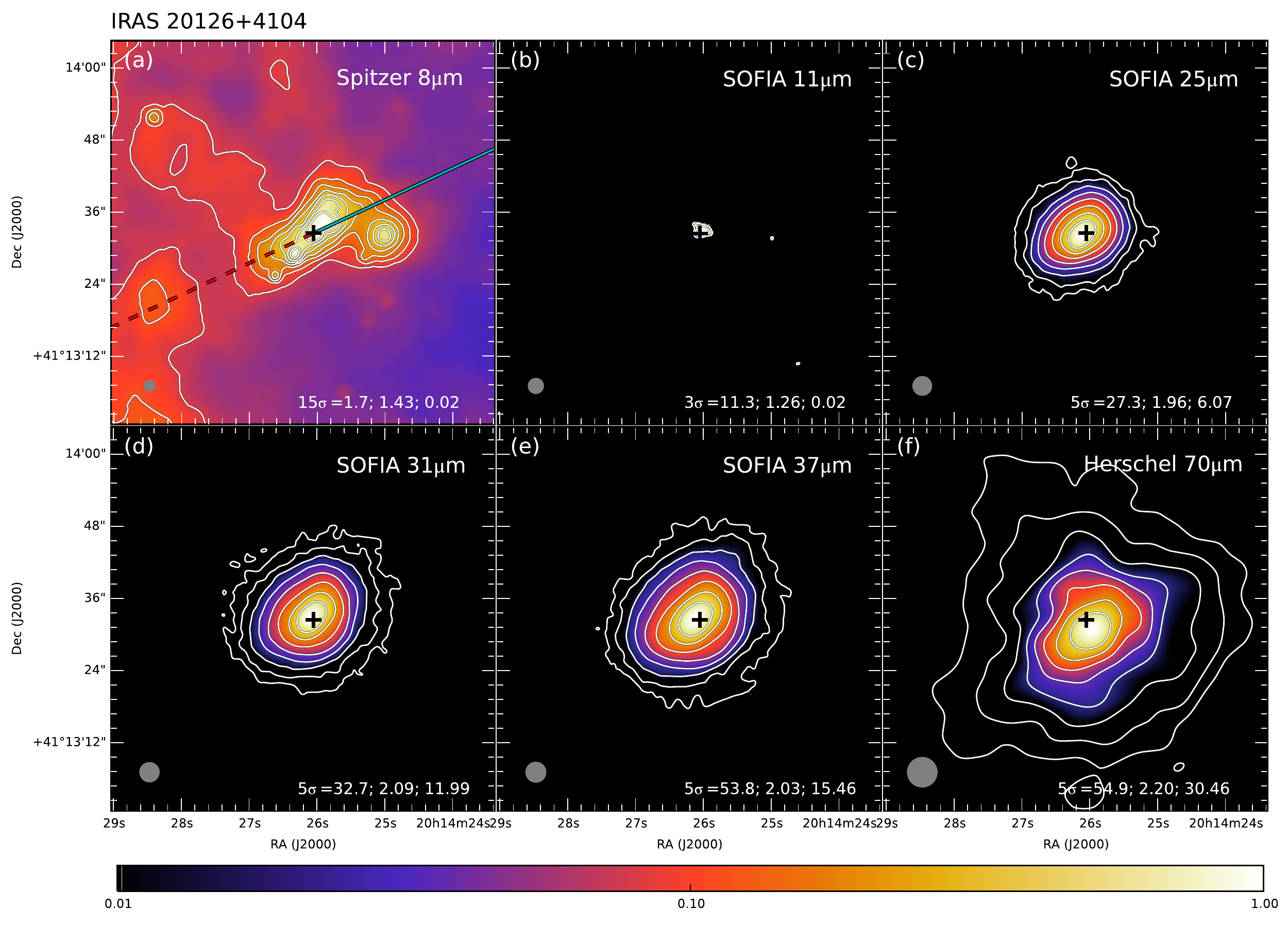}
\caption{
Multiwavelength images of IRAS 20126+4104, following the format of
Fig.~\ref{fig:AFGL4029}. The nominal location of the protostar, derived
from the model fit to the proper motions of the water masers from
Moscadelli et al. (2011), is shown as a large cross in all panels at
R.A.(J2000) = 20$^h$14$^m$26$\fs$05, decl.(J2000) =
$+$41$\arcdeg$13$\arcmin$32$\farcs$48. The outflow axis angle and the
blueshifted outflow direction are given by the HCO$^+$ observations
of Cesaroni et al. (1999).\label{fig:IRAS20126}}
\end{figure*}

At a distance of 1.64 kpc (Moscadelli et al. 2011) in the Cygnus-X
star-forming region, IRAS~20116+4104, along with G35.20-0.74, could be
considered a prototypical example of an MYSO with an outflow, and
consequently, there have been numerous studies directed toward this
object. Observations suggest a luminosity of
$1.3\times10^4\:L_{\odot}$ with a central protostar having an
estimated mass of 7 to $12\:M_{\odot}$ (Cesaroni et al. 1997; Keto \&
Zhang 2010; Johnston et al. 2011; Chen et al. 2016). This source is
surrounded by a resolved accretion disk, believed to be undergoing
Keplerian rotation (Cesaroni et al. 1997; 1999; 2005) at a position
angle of $\sim$53$^{\circ}$. Though IRAS~20116+4104 appears to be an
MYSO, it might be too young to have produced a UC~\ion{H}{2} region;
radio continuum emission observations at centimeter wavelengths show
that the emission components near the center of the outflow are
consistent with free-free emission from ionized gas in an outflow. The
location of the driving source of the outflow was determined through
proper motion studies of water masers, which seem to be moving away
from a common location (Moscadelli et al. 2011). This location is
coincident with the center of the accretion disk as delineated by
CH$_3$CN(12-11) emission from Cesaroni et al. (1999).

IRAS~20126+4104 has a well-collimated bipolar molecular outflow
oriented at an angle roughly perpendicular to the disk
(P.A.$\sim$115$^{\circ}$) with an inclination angle of the outflow
axis to the plane of the sky of only $\sim$10$^{\circ}$ (Zhang et
al. 1999; Hofner et al. 2007; Su et al. 2007; Moscadelli et el. 2011;
see also Cesaroni et al. 2013). De Buizer (2007) made the first
suggestion that the extended MIR emission observed toward this source
might be related to the outflow.

At wavelengths greater than 19\,$\mu$m, SOFIA data
(Figure~\ref{fig:IRAS20126}) show an elongated morphology at an angle
(P.A.$\sim$125$^{\circ}$) similar to that of the outflow (the
11\,$\mu$m SOFIA observation is a shallow integration that
only barely detects the peak emission from the source). Even the
\textit{Herschel} 70\,$\mu$m data show an elongation along this
outflow axis direction. The location of the driving source from
Moscadelli et al. (2011) is coincident with the MIR/FIR peak (to
within astrometric accuracy), and the amount of extended emission seen
to the NW of this peak is comparable to that seen to the SE. This
symmetry may be expected since the outflow is oriented almost in the
plane of the sky, and consequently there should be little bias of
emission from just the blueshifted lobe.


\subsubsection{Cepheus~A}

\begin{figure*}
\epsscale{1.15}
\plotone{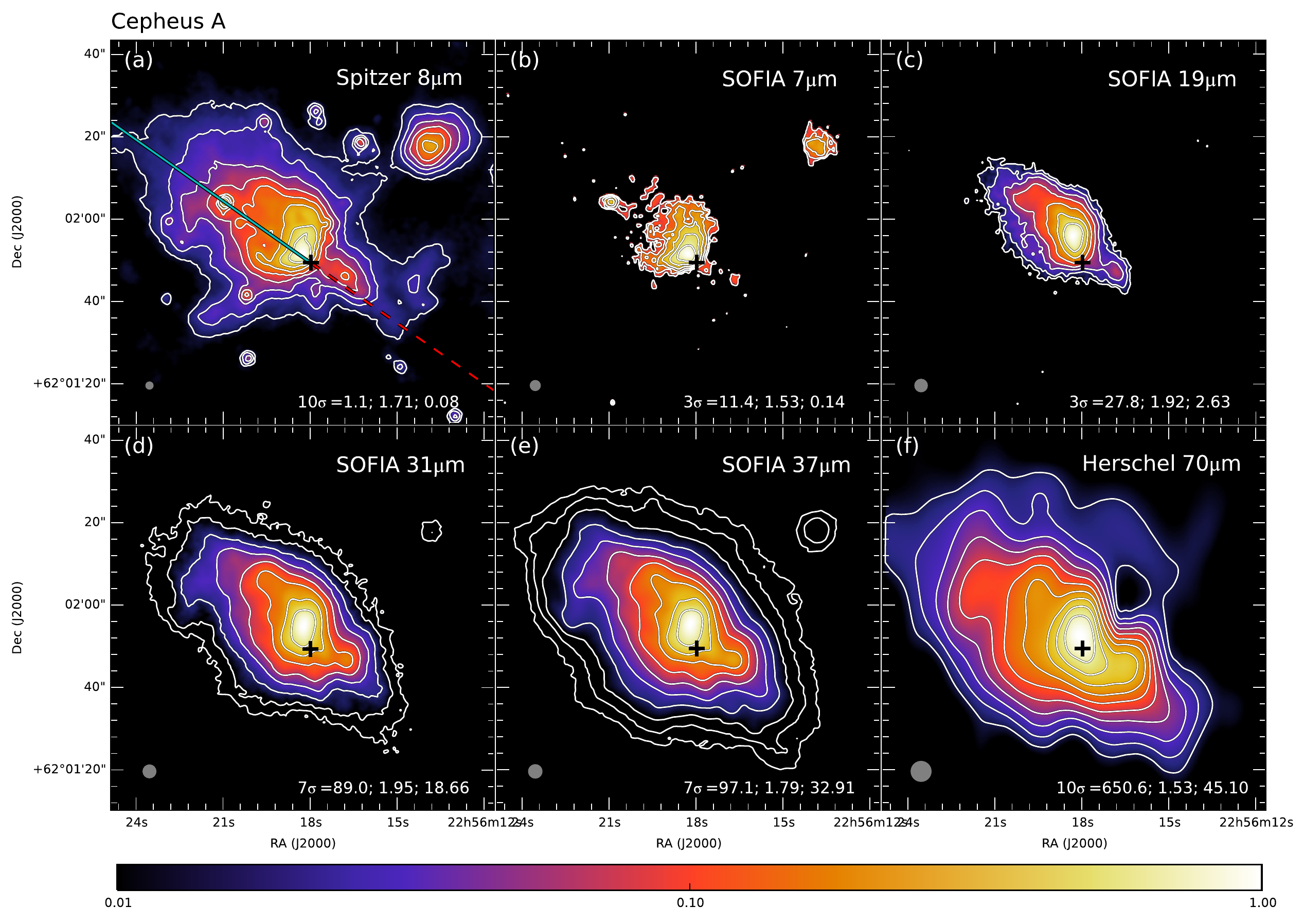}
\caption{
Multiwavelength images of Cepheus A, following the format of
Fig.~\ref{fig:AFGL4029}. The cross in each panel shows the
location of the radio continuum source HW~2 at R.A.(J2000) =
22$^h$56$^m$17$\fs$98, decl.(J2000) =
$+$62$\arcdeg$01$\arcmin$49$\farcs$39. The outflow axis angle and the
blueshifted outflow direction are given by the HCO$^+$ observations
of G\'omez et al. (1999).\label{fig:CepA}}
\end{figure*}

Cep A contains a massive bipolar molecular outflow primarily aligned
east-west that was initially identified by Rodriguez et al. (1980);
however, at higher spatial resolutions, the outflow morphology is quite
complex. The central $\sim$2$\arcmin$ of the outflow appears to be
dominated by components aligned NE-SW (Bally \& Lane 1990; Torrelles
et al. 1993; Narayanan \& Walker 1996; G{\'o}mez et al. 1999; Zapata
et al. 2013). This central region contains a compact, extremely
high-velocity CO outflow (Narayanan \& Walker 1996) with an axis at a
position angle of $\sim$50$^{\circ}$ that is believed to trace a
younger component than the rest of the outflow (Cunningham et
al. 2009). This central outflow component appears to have an axis
close to the plane of the sky but with blueshifted emission to the NE
(G\'omez et al. 1999; Zapata et al. 2013). At NIR wavelengths, the region
displays an extremely bright reflection nebula (Cunningham et
al. 2009), almost wholly contained within this blueshifted outflow
cavity.

At the center of this outflow is a cluster of radio sources, and there
is confusion as to which source(s) might be driving the outflow(s)
(Zapata et al. 2013). One of the main candidates for driving the
outflow, and the brightest radio continuum source in the region, is
HW~2 (Hughes \& Wouterloot 1984). It has a luminosity of about
$10^4\:L_{\odot}$ (Garay et al. 1996), suggesting it is a B0.5 star
approaching $20\:M_{\odot}$, given a distance to the source of 700~pc
based on parallax measurements of 12 GHz methanol masers in the region
(Moscadelli et al. 2009) and of radio source HW~9 (Dzib et
al. 2011). HW~2 has not been detected at NIR wavelengths (Casement \&
McLean 1996; Jones et al. 2004; Cunningham et al. 2009), nor in the
MIR (De Buizer et al. 2005; de Wit et al. 2009; also Cunningham et
al. 2009; however, the absolute astrometry of their MIR images, and
hence the placement of radio sources with respect to the MIR sources,
appears to be off by over 6$\arcsec$).

The estimated extinction to the region around HW~2 is
$A_V\sim$300--1000 mag (Goetz et al. 1998; Cunningham et
al. 2009), and therefore it is not surprising it is not directly
detected in the NIR, MIR, or in our SOFIA data
(Figure~\ref{fig:CepA}). However, it does appear that the contour peak
shifts toward this location in the 70\,$\mu$m \textit{Herschel} data
(Figure~\ref{fig:CepA}(f)).

At 7\,$\mu$m the emission seen by SOFIA corresponds well to
the NIR reflection nebula and blueshifted outflow cavity. As one goes
to longer SOFIA wavelengths, we begin to see increasingly
brighter emission to the SW, which corresponds to the direction of the
redshifted outflow. We suggest that we are beginning to penetrate the
higher extinction toward this region and the emission we are seeing
at wavelengths $>$30\,$\mu$m is coming from the redshifted outflow
cavity.

\subsubsection{NGC~7538 IRS~9}\label{S:NGC7539IRS9}

\begin{figure*}
\epsscale{1.15}
\plotone{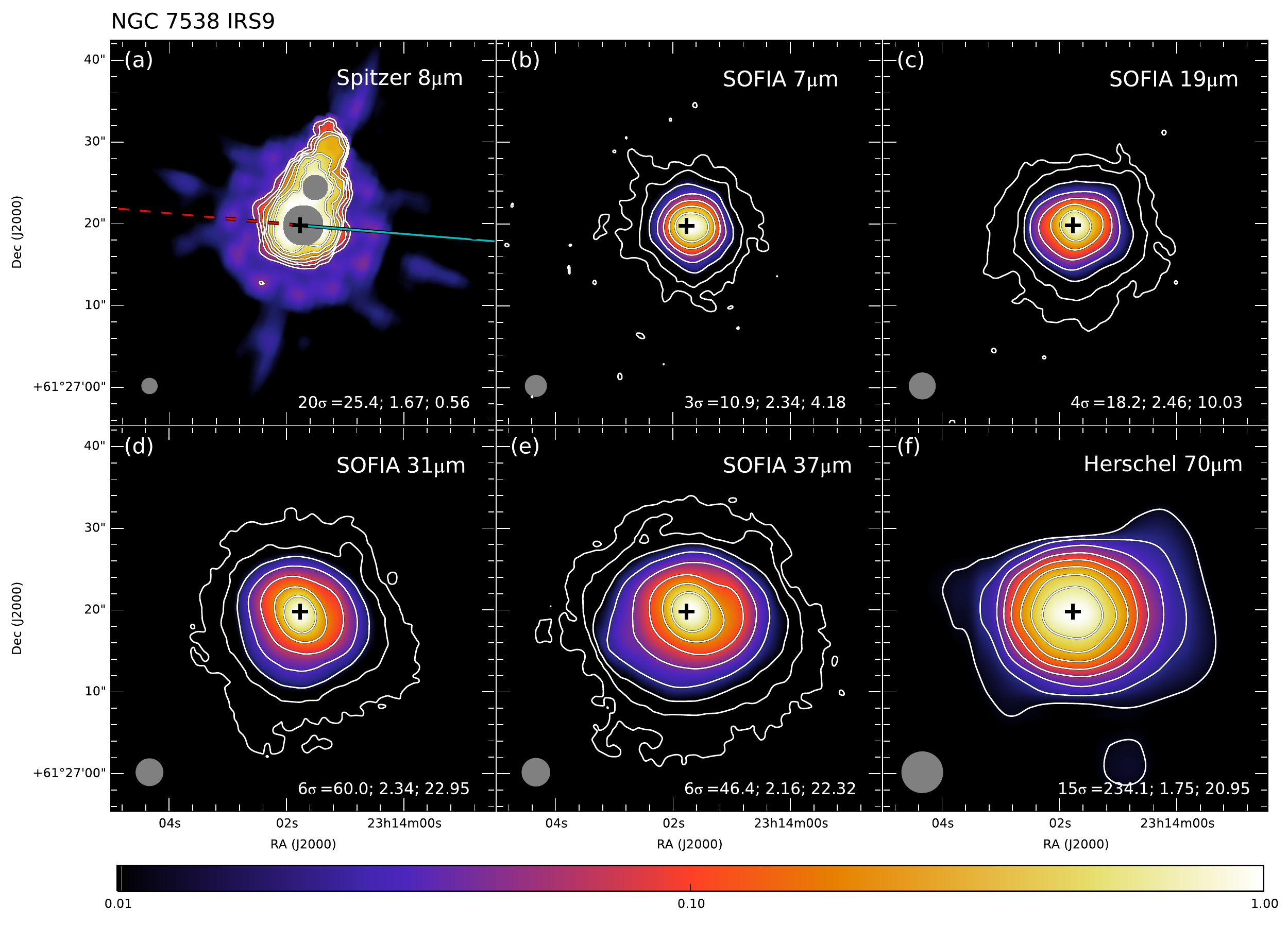}
\caption{
Multiwavelength images of NGC 7538 IRS9, following the format of
Fig.~\ref{fig:AFGL4029}. The gray areas in panel (a) are where the
source has saturated in the IRAC image. The extension to the northwest
in panel (a) is a ghost, and not a real structure. The location of the
3.6~cm radio continuum peak from Sandell et al. (2005) is shown as a
large cross in all panels at R.A.(J2000) = 23$^h$14$^m$01$\fs$77,
decl.(J2000) = $+$61$\arcdeg$27$\arcmin$19$\farcs$8. The outflow axis
angle and the blueshifted outflow direction are given by the HCO$^+$
observations of Sandell et al. (2005). \label{fig:NGC7538IRS9}}
\end{figure*}

NGC~7538 is an optically visible \ion{H}{2} region (Fich \& Blitz
1984) located at a distance of 2.65~kpc, as determined from
trigonometric parallax measurements (Moscadelli et al. 2009). Infrared
observations of this region by Wynn-Williams et al. (1974) and Werner
et al. (1979) led to the identification of multiple discrete sources
in the vicinity of the optical nebula, which were named IRS~1 through
11. The source IRS~9 lies $\sim$2$\arcmin$ to the SE of the prominent
and well-studied IRS~1 region. It powers its own reflection nebula,
and has a total luminosity of about $3.5\times10^4\:L_{\odot}$
(Sandell et al. 2005, corrected to the distance from Moscadelli et
al. 2009), which is the equivalent of a B0.5 ZAMS star.

Though IRS~9 has the luminosity of a typical MYSO, it has very weak
radio continuum emission. Sandell et al. (2005) found that the object
has a flat radio spectrum consistent with free-free emission from a
collimated, ionized jet. They also disentangled the rather complex
structures seen in various outflow tracers into distinct outflows from
three different sources, suggesting a cluster associated with
IRS~9. The outflow associated most closely with the position of IRS~9
itself was measured to have a very high velocity (Mitchell \& Hasegawa
1991), leading to the suggestion that we might be observing the system
nearly face-on (Barentine \& Lacy 2012). The high spatial resolution
($\sim$6$\arcsec$) HCO$^+$ maps of Sandell et al. (2005) show that
IRS~9 indeed drives a bipolar, extremely high-velocity outflow
approximately oriented E-W (P.A.$\sim$85$^{\circ}$) that is inclined
by only $\sim$20$^{\circ}$ to the line of sight. Given this
orientation, the outflow lobes seen in HCO$^+$ do not extend very far
from IRS~9 in projection ($\sim$14$\arcsec$), but the blueshifted
outflow lobe is clearly to the west of IRS~9, and the redshifted
outflow lobe to the east (Figure~\ref{fig:NGC7538IRS9}a). We note here
that the best-fitting ZT and Robitaille et al. radiative transfer
models for this system (presented below), based solely on SED fitting,
have viewing angles of about 20$^\circ$ to the outflow axis, very
similar to the above estimates based on outflow observations.

Our SOFIA data for this source look rather point like at
7\,$\mu$m; however, beginning at 19\,$\mu$m the source begins to show
signs of being elongated in an E-W orientation, similar to the outflow
axis (Figure~\ref{fig:NGC7538IRS9}). The {\it Herschel} 70\,$\mu$m
data also show a more prominent east-west elongation with the a larger
extension to the west in the direction of the blueshifted outflow
cavity.

\subsection{General Results from the SOFIA Imaging}

\begin{figure*}
\epsscale{1.15}
\plotone{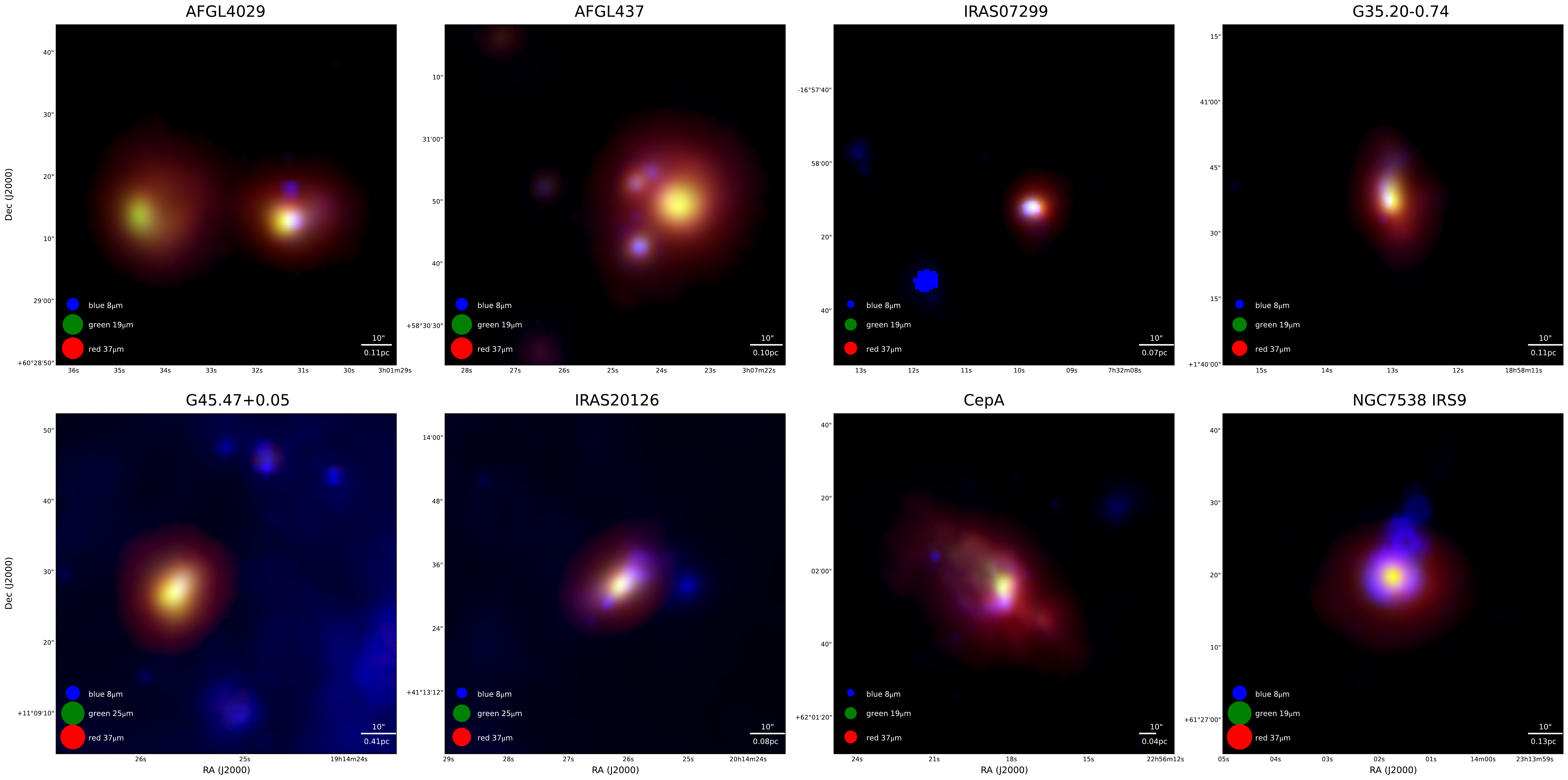}
\caption{
Gallery of RGB 
images of the eight protostellar sources, as labeled. The legend
shows the wavelengths used and the beam sizes at these
wavelengths. SOFIA-FORCAST~37~$\rm \mu m$ is always shown in red, and
{\it Spitzer}-IRAC 8~$\mu m$ is always shown in blue (note this
occasionally saturates in the brightest parts of some sources: see
previous individual source images). Green usually shows
SOFIA-FORCAST~19~$\rm \mu m$, except for G45.47+0.05 and IRAS 20126,
where it displays FORCAST~25~$\rm \mu m$.
\label{fig:rgb}}
\end{figure*}

In addition to the monochromatic images presented above, we also
construct three-color images of all the sources, presented together in
Figure~\ref{fig:rgb}. The three-color images reveal color gradients
across the sources: i.e., the more extincted, far-facing outflow
cavities appear redder, with this morphology particular clear in the
cases of G35.20-0.74 and Cep A. Note, however, that these RGB images
have different beam sizes for the different colors (especially blue),
with the effect being to tend to give small sources an extended red
halo.

G35.20-0.74 was the first source observed for this survey, and it has
been the subject of its own paper (Zhang et al. 2013b) describing how
the outflow from this massive protostar is likely to directly
influence the morphology we see at infrared wavelengths. The
hypothesis is that massive stars form in dense cores, with extinctions
of $A_V\gtrsim$ hundreds of magnitudes along the line of sight to the
central protostar. Outflows are driven by accretion and can
effectively clear out material surrounding the core along the outflow
axis direction, significantly decreasing extinction in those
directions. Thus, radiation readily leaves via these cavities, and if
the orientation to our line of sight is favorable, we can detect more
intense and shorter wavelength infrared emission from these
sources. Blueshifted outflow cavities appear brighter. However, as one
observes at longer wavelengths, it becomes possible to see emission
from the redshifted outflow cavities. The previous subsection
discussed the observational evidence that indicates that each of the
regions in our sample contains a high- or intermediate-mass protostar
driving an outflow. How wide-spread is the evidence in our sample that
the MIR morphologies are influenced by the presence of these outflow
cavities?

Of the eight sources in our sample, only AFGL~437 
does not show clear signs of extended MIR/FIR emission. Of the
remaining seven sources, we can conclude that six are extended in
their MIR/FIR emission at a position angle comparable to the
orientation of their outflow axes. The only exception is
IRAS~07299-1651, and this is only excluded because no outflow maps
exist for this source. However, since it displays a behavior in
morphology as a function of wavelength similar to the rest of the
sources, we predict that an outflow is present at a position angle of
$\sim$300$^{\circ}$, with a blueshifted lobe to the SE. For two of the
sources in the sample, it appears that their MIR/FIR emission is
extended only to one side of the central stellar source: AFGL~4029 and
G45.47+0.05. In both cases, this emission is on the blueshifted
side. Three sources appear to be extended to one side at shorter
wavelengths and more symmetrically extended at longer wavelengths:
G35.20-0.74, IRAS~20126+4104, and Cepheus A. In all three cases, the
emission at shorter wavelengths comes predominantly from the
blueshifted side of the outflow.
The remaining source is NGC~7538~IRS~9, for which, perhaps because of
an almost pole-on outflow orientation, we only see modest amounts of
extended MIR/FIR emission. However, the little MIR/FIR extension that
is seen is at the angle of the projected outflow axis. Somewhat
surprising, however, is that the elongated morphologies seen at
7--40\,$\mu$m are also present in most cases in the {\it Herschel}
70\,$\mu$m images, showing that outflows can impact protostellar
appearance even at such long FIR wavelengths.

Thus, the first eight sources of the SOMA Star Formation survey give
strong support to the hypothesis that MIR to FIR morphologies of high-
and intermediate-mass protostars are shaped by their outflow
cavities. Bipolar, oppositely directed outflows are a generic
prediction of Core Accretion models. The presence of dense core
envelope gas near the protostar will tend to extinct shorter
wavelength light to a greater degree so that the emission peaks at
these wavelengths appear displaced away from the protostar toward the
blueshifted, near-facing side of the outflow. This qualitative
prediction again appears to be confirmed by our survey results. MIR to
FIR morphologies thus give important information about how massive
protostars are forming, especially the orientation and structure of
their outflow cavities and the presence of dense core envelopes. In
the following section, we use the SOFIA and other data to make
more quantitative assessments of the properties of these protostars.

\subsection{Results of SED Model Fitting}


Here we focus on simple SED model fits to the sample, deferring the
fitting of image flux profiles to a future paper. We will compare the
results derived from the ZT model grid with those from the Robitaille
et al. grid.

\subsubsection{The SEDs}

\begin{figure*}
\epsscale{1.15}
\plotone{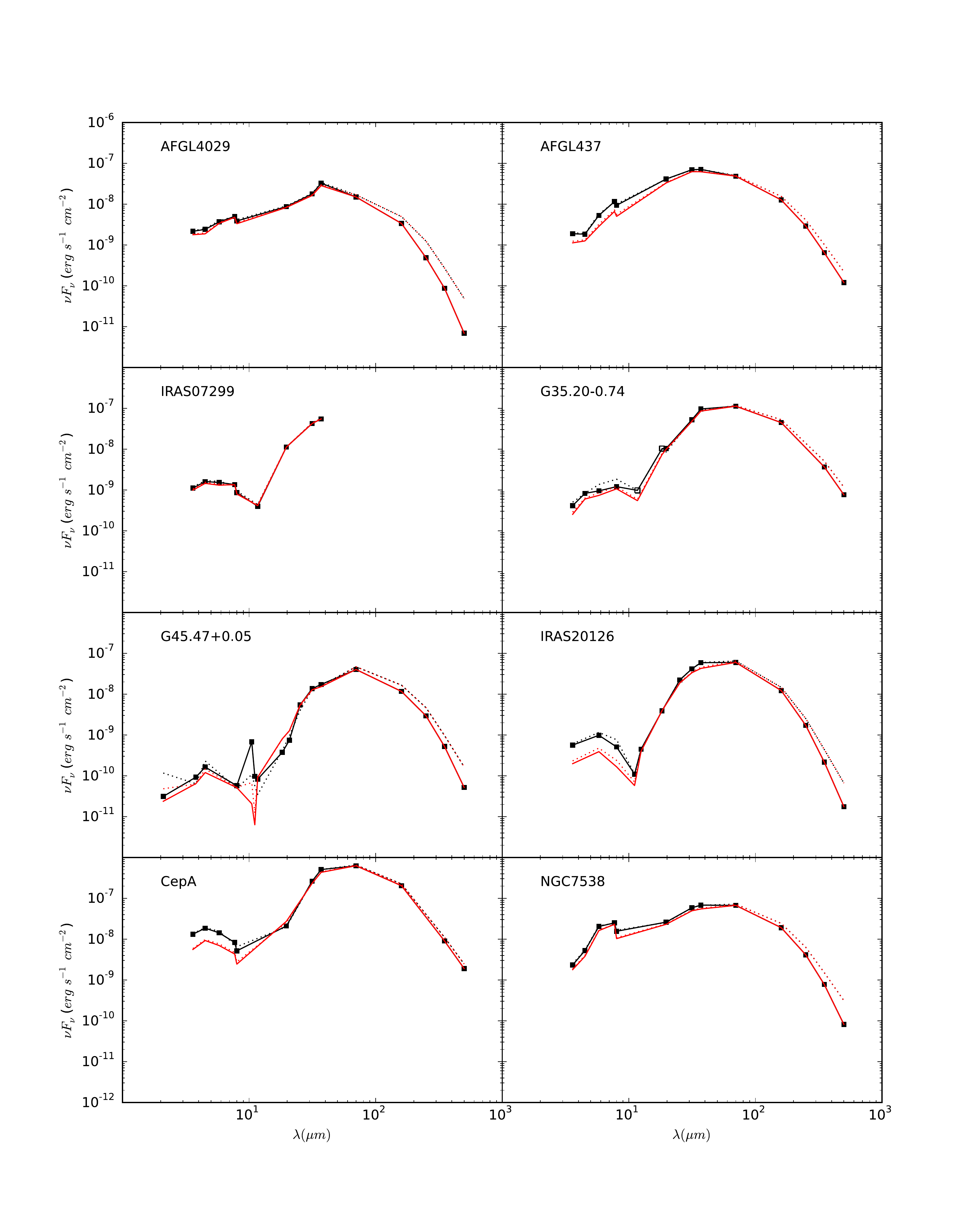}
\caption{
SEDs of the first eight sources of the SOMA Survey. Total fluxes with
no background subtraction applied are shown by dotted lines. The fixed
aperture case is the black dotted line; the variable aperture (at
$<70\:{\rm \mu m}$) case is the red dotted line. The
background-subtracted SEDs are shown by solid lines: black for fixed
aperture (the fiducial case); red for variable aperture. Black solid
squares indicate the actual measured values that sample the fiducial
SED. Note that the open squares in the Gemini data of G35.20-0.74 are
values where no background subtraction could be done given the limited
field of view of the observations.
\label{fig:SEDs}}
\end{figure*}

Figure~\ref{fig:SEDs} shows the SEDs of the eight sources that
have been discussed in this paper. The figure illustrates the effects
of using fixed or variable apertures, as well as the effect of
background subtraction. Our fiducial method is that with fixed
aperture and with background subtraction carried out. This tends to
have moderately larger fluxes at shorter wavelengths than the variable
aperture SED. However, the $\lesssim8\:{\rm \mu m}$ flux is in any case
treated as an upper limit in the SED model fitting, given the
difficulties of modeling emission from PAHs and transiently heated
small grains. 
Apart from IRAS~07299-1651, which lacks {\it Herschel} data, all of
the SEDs are well characterized: in particular, the peaks are well
covered by the combination of SOFIA-FORCAST and {\it Herschel}-PACS
and SPIRE data.

We note that in the case of G35.20-0.74, our derived fiducial SED
differs modestly ($\lesssim20\%$) from that estimated by Zhang et
al. (2013b). These differences are due to our use of a fixed aperture
size and geometry. Also, our SED now replaces {\it IRAS} fluxes with
those measured by {\it Herschel}.

\subsubsection{ZT Model Fitting Results}

\begin{figure*}
\epsscale{1.0}
\plotone{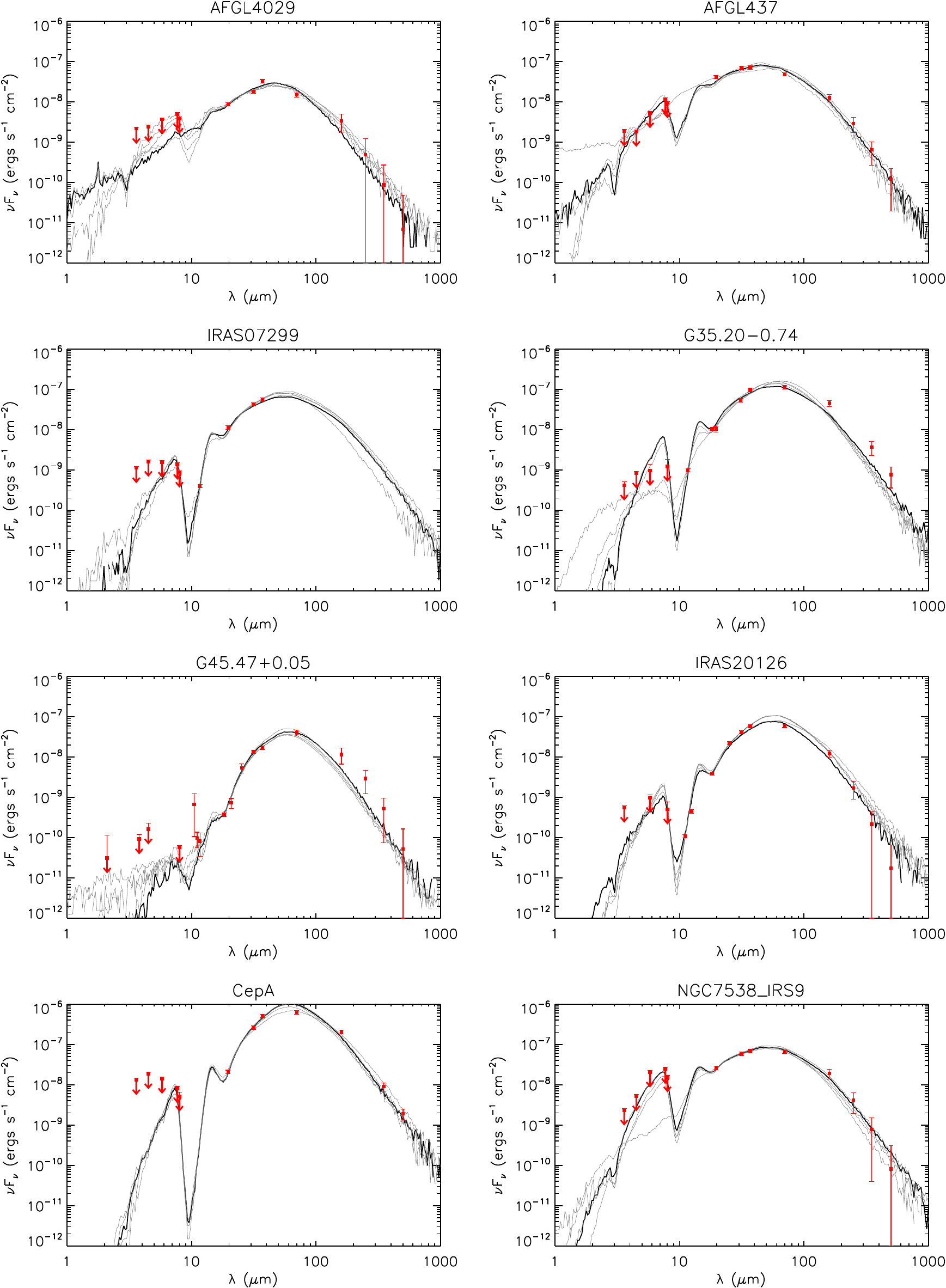}
\caption{
Protostar model fitting to the fixed aperture, background-subtracted
SED data using the ZT model grid. For each source, the best-fit model
is shown with a solid black line and the next four best models are
shown with solid gray lines. Flux values are those from
Table~\ref{tab:flux}. Note that the data at $\lesssim8\:{\rm \mu m}$
are treated as upper limits (see text). The resulting model parameter
results are listed in Table~\ref{tab:models}.\label{fig:SEDsZT}}
\end{figure*}

Figure~\ref{fig:SEDsZT} shows the results of fitting the ZT
protostellar radiative transfer models to the fixed aperture,
background-subtracted SEDs. Note that the data at $\lesssim8\:{\rm \mu
  m}$ are considered to be upper limits given that PAH emission and
transiently heated small grain emission are not well treated in the
models.

\begin{sidewaystable*}
\centering
\setlength{\tabcolsep}{3pt}
\renewcommand{\arraystretch}{0.9}
\vspace{3.5in}
\begin{deluxetable*}{c|ccccccccccc|cccccccccc}
\tabletypesize{\scriptsize}
\tablecaption{Parameters of the Five Best-fitted Models of Zhang \& Tan and Robitaille et al. models\label{tab:models}}
\tablewidth{16pt}
\tablehead{
\colhead{} & \multicolumn{11}{c}{Zhang \& Tan Models} & \multicolumn{10}{c}{Robitaille et al. Models} \\
\colhead{Source} &\colhead{$\chi^{2}$} & \colhead{$M_{c}$} & \colhead{$\Sigma_{\rm cl}$} & \colhead{$R_{c}$} &\colhead{$m_{*}$} & \colhead{$\theta_{\rm view}$} &\colhead{$A_{V}$} & \colhead{$M_{\rm env}$} &\colhead{$\theta_{w,\rm esc}$} & \colhead{$\dot {M}_{\rm disk}$} & \colhead{$L_{\rm bol}$} &\colhead{$\chi^{2}$} &\colhead{$m_{*}$} & \colhead{$\theta_{\rm view}$} &\colhead{$A_{V}$} & \colhead{$M_{\rm env}$} & \colhead{$R_{\rm env}$} &\colhead{$\theta_{w,\rm esc}$} & \colhead{$\dot {M}_{\rm env}$} & \colhead{$\dot {M}_{\rm disk}$} & \colhead{$L_{\rm bol}$} \\
\colhead{} & \colhead{} & \colhead{($M_\odot$)} & \colhead{(g $\rm cm^{-2}$)} & \colhead{(pc) ($\arcsec$)} & \colhead{($M_{\odot}$)} & \colhead{($^\circ$)} & \colhead{(mag)} & \colhead{($M_{\odot}$)} & \colhead{($^\circ$)} &\colhead{($M_{\odot}$/yr)} & \colhead{($L_{\odot}$)} & \colhead{} & \colhead{($M_{\odot}$)} & \colhead{($^\circ$)} & \colhead{(mag)} & \colhead{($M_{\odot}$)} & \colhead{(pc) ($\arcsec$)} & \colhead{($^\circ$)} & \colhead{($M_{\odot}$/yr)} & \colhead{($M_{\odot}$/yr)} & \colhead{($L_{\odot}$)} 
}
\startdata
AFGL 4029
& 1.08 & 30 & 1.0 & 0.04 (4) & 12 & 65 & 2.0 & 6 & 53 & 1.9(-4) & 4.1(4) & 1.79 & 12 & 57 & 62.3 & 58 & 0.34 (36) & 36 & 2.0(-4) & 5.8(-8) & 1.0(4) \\
$d$ = 2.0 kpc & 1.32 & 160 & 0.3 & 0.17 (17) & 48 & 86 & 23.2 & 14 & 77 & 1.1(-4) & 3.4(5) & 1.79 & 12 & 87 & 46.9 & 83 & 0.48 (50) & 34 & 1.7(-4) & 1.8(-6) & 1.1(4) \\
$R_{\rm ap}$ = 11.2\arcsec & 1.33 & 80 & 0.1 & 0.21 (21) & 12 & 77 & 2.0 & 47 & 40 & 5.4(-5) & 1.6(4) & 1.80 & 11 & 18 & 45.8 & 79 & 0.31 (32) & 44 & 3.9(-4) & 8.7(-9) & 9.1(3) \\
& 1.52 & 60 & 0.3 & 0.10 (10) & 16 & 62 & 4.0 & 19 & 56 & 1.1(-4) & 3.6(4) & 1.81 & 13 & 63 & 56.3 & 69 & 0.48 (50) & 42 & 1.7(-4) & 4.4(-7) & 1.2(4) \\
& 1.53 & 50 & 0.3 & 0.09 (10) & 12 & 55 & 1.0 & 22 & 46 & 1.0(-4) & 2.4(4) & 1.81 & 14 & 49 & 53.0 & 66 & 0.48 (49) & 20 & 1.1(-4) & 1.6(-8) & 8.8(3) \\
\hline\noalign{\smallskip}
AFGL 437
& 0.91 & 160 & 0.1 & 0.29 (30) & 16 & 58 & 0.0 & 116 & 32 & 8.1(-5) & 3.3(4) & 0.79 & 15 & 87 & 15.7 & 124 & 0.48 (50) & 35 & 2.9(-4) & 9.7(-6) & 2.3(4) \\
$d$ = 2.0 kpc & 1.48 & 160 & 0.1 & 0.29 (30) & 24 & 86 & 15.2 & 87 & 45 & 8.5(-5) & 7.8(4) & 0.83 & 15 & 81 & 16.3 & 124 & 0.48 (50) & 35 & 2.9(-4) & 9.7(-6) & 2.3(4) \\
$R_{\rm ap}$ = 32.0\arcsec & 1.55 & 50 & 3.2 & 0.03 (3) & 8 & 29 & 0.0 & 35 & 25 & 6.0(-4) & 1.7(4) & 1.05 & 16 & 76 & 12.9 & 97 & 0.48 (50) & 17 & 1.8(-4) & 2.9(-4) & 2.0(4) \\
& 2.02 & 160 & 0.1 & 0.29 (30) & 32 & 89 & 23.2 & 55 & 59 & 7.6(-5) & 1.5(5) & 1.07 & 14 & 81 & 10.0 & 141 & 0.48 (50) & 30 & 2.9(-4) & 3.0(-7) & 1.9(4) \\
& 2.22 & 200 & 0.1 & 0.33 (34) & 12 & 34 & 0.0 & 174 & 20 & 8.0(-5) & 2.0(4) & 1.07 & 16 & 87 & 10.0 & 161 & 0.48 (50) & 24 & 3.1(-4) & 1.8(-7) & 2.3(4) \\
\hline\noalign{\smallskip}
IRAS 07299
& 0.54 & 240 & 0.1 & 0.36 (44) & 8 & 86 & 5.1 & 226 & 13 & 7.1(-5) & 1.1(4) & 1.49 & 19 & 81 & 27.6 & 35 & 0.14 (17) & 5 & 4.4(-4) & ... & 1.0(4) \\
$d$ = 1.68 kpc & 0.62 & 80 & 0.3 & 0.12 (14) & 16 & 89 & 11.1 & 42 & 42 & 1.5(-4) & 4.2(4) & 1.52 & 19 & 76 & 28.2 & 35 & 0.14 (17) & 5 & 4.4(-4) & ... & 1.0(4) \\
$R_{\rm ap}$ = 7.7\arcsec & 0.86 & 240 & 0.1 & 0.36 (44) & 12 & 89 & 49.5 & 211 & 19 & 8.5(-5) & 2.0(4) & 1.55 & 19 & 87 & 27.9 & 35 & 0.14 (17) & 5 & 4.4(-4) & ... & 1.0(4) \\
& 1.14 & 200 & 0.1 & 0.33 (40) & 12 & 86 & 51.5 & 174 & 20 & 8.0(-5) & 2.0(4) & 1.58 & 18 & 87 & 17.0 & 70 & 0.23 (29) & 7 & 3.8(-4) & ... & 8.7(3) \\
& 1.24 & 400 & 0.1 & 0.47 (57) & 8 & 62 & 0.0 & 386 & 10 & 8.2(-5) & 1.0(4) & 1.60 & 19 & 63 & 29.7 & 35 & 0.14 (17) & 5 & 4.4(-4) & ... & 1.0(4) \\
\hline\noalign{\smallskip}
G35.20-0.74
& 2.63 & 480 & 0.1 & 0.51 (48) & 16 & 48 & 40.4 & 440 & 15 & 1.2(-4) & 3.8(4) & 2.26 & 20 & 87 & 20.7 & 597 & 0.48 (45) & 34 & 1.6(-3) & 2.8(-7) & 4.7(4) \\
$d$ = 2.2 kpc & 2.64 & 100 & 3.2 & 0.04 (4) & 12 & 29 & 70.7 & 77 & 20 & 9.4(-4) & 5.2(4) & 2.40 & 20 & 81 & 24.1 & 597 & 0.48 (45) & 34 & 1.6(-3) & 2.8(-7) & 4.7(4) \\
$R_{\rm ap}$ = 32.0\arcsec & 2.76 & 320 & 0.1 & 0.42 (39) & 24 & 68 & 81.8 & 256 & 27 & 1.2(-4) & 8.4(4) & 2.49 & 20 & 76 & 33.0 & 597 & 0.48 (45) & 34 & 1.6(-3) & 2.8(-7) & 4.7(4) \\
& 2.76 & 80 & 3.2 & 0.04 (3) & 12 & 39 & 15.2 & 58 & 22 & 8.4(-4) & 5.0(4) & 2.54 & 19 & 70 & 16.4 & 679 & 0.48 (45) & 27 & 1.5(-3) & 2.6(-7) & 4.3(4) \\
& 2.77 & 200 & 0.3 & 0.19 (17) & 12 & 22 & 43.4 & 173 & 17 & 1.9(-4) & 4.0(4) & 2.70 & 18 & 76 & 16.8 & 560 & 0.48 (45) & 29 & 1.2(-3) & 3.9(-6) & 3.6(4) \\
\hline\noalign{\smallskip}
G45.47+0.05
& 1.21 & 200 & 3.2 & 0.06 (1) & 32 & 86 & 63.6 & 140 & 25 & 1.7(-3) & 4.6(5) & 3.36 & 31 & 57 & 11.1 & 1562 & 0.48 (12) & 20 & 4.1(-3) & ... & 1.4(5) \\
$d$ = 8.4 kpc & 1.34 & 320 & 1.0 & 0.13 (3) & 48 & 89 & 46.5 & 200 & 35 & 9.3(-4) & 5.1(5) & 3.67 & 34 & 63 & 10.0 & 1725 & 0.48 (12) & 19 & 4.7(-3) & ... & 1.7(5) \\
$R_{\rm ap}$ = 14.4\arcsec & 1.57 & 320 & 1.0 & 0.13 (3) & 32 & 68 & 15.2 & 252 & 24 & 8.2(-4) & 2.7(5) & 3.94 & 29 & 70 & 15.2 & 967 & 0.48 (12) & 17 & 2.4(-3) & ... & 1.2(5) \\
& 1.62 & 240 & 1.0 & 0.11 (3) & 32 & 86 & 1.0 & 170 & 30 & 7.2(-4) & 2.6(5) & 3.98 & 29 & 81 & 10.0 & 967 & 0.48 (12) & 17 & 2.4(-3) & ... & 1.2(5) \\
& 1.75 & 240 & 1.0 & 0.11 (3) & 24 & 55 & 0.0 & 192 & 23 & 6.6(-4) & 1.7(5) & 3.99 & 34 & 81 & 37.7 & 1008 & 0.48 (12) & 24 & 2.9(-3) & ... & 1.7(5) \\
\hline\noalign{\smallskip}
IRAS 20126
& 1.82 & 80 & 0.3 & 0.12 (15) & 16 & 74 & 37.4 & 42 & 42 & 1.5(-4) & 4.2(4) & 1.10 & 18 & 76 & 92.4 & 230 & 0.48 (61) & 17 & 4.4(-4) & 5.7(-7) & 2.3(4) \\
$d$ = 1.64 kpc & 2.07 & 120 & 0.3 & 0.14 (18) & 24 & 74 & 69.7 & 57 & 47 & 1.8(-4) & 9.3(4) & 1.10 & 18 & 70 & 96.7 & 230 & 0.48 (61) & 17 & 4.4(-4) & 5.7(-7) & 2.3(4) \\
$R_{\rm ap}$ = 12.8\arcsec & 2.32 & 80 & 0.3 & 0.12 (15) & 12 & 44 & 73.7 & 53 & 31 & 1.4(-4) & 3.4(4) & 1.11 & 18 & 87 & 89.9 & 230 & 0.48 (61) & 17 & 4.4(-4) & 5.7(-7) & 2.3(4) \\
& 2.33 & 200 & 0.1 & 0.33 (41) & 12 & 86 & 65.7 & 174 & 20 & 8.0(-5) & 2.0(4) & 1.14 & 18 & 81 & 90.9 & 230 & 0.48 (61) & 17 & 4.4(-4) & 5.7(-7) & 2.3(4) \\
& 2.39 & 100 & 0.3 & 0.13 (16) & 16 & 51 & 66.7 & 61 & 36 & 1.6(-4) & 4.5(4) & 1.26 & 18 & 70 & 107.9 & 107 & 0.35 (44) & 13 & 3.2(-4) & 1.0(-5) & 2.4(4) \\
\hline\noalign{\smallskip}
Cep A
& 2.23 & 160 & 0.3 & 0.17 (49) & 16 & 44 & 94.9 & 125 & 26 & 2.0(-4) & 5.0(4) & 1.47 & 19 & 57 & 57.1 & 723 & 0.48 (143) & 17 & 1.4(-3) & 5.9(-5) & 2.8(4) \\
$d$ = 0.70 kpc & 2.30 & 160 & 0.3 & 0.17 (49) & 12 & 29 & 100.0 & 135 & 20 & 1.8(-4) & 3.8(4) & 1.47 & 19 & 63 & 47.1 & 723 & 0.48 (143) & 17 & 1.4(-3) & 5.9(-5) & 2.8(4) \\
$R_{\rm ap}$ = 48.0\arcsec & 2.32 & 480 & 0.1 & 0.51 (150) & 12 & 83 & 85.9 & 460 & 12 & 1.1(-4) & 2.4(4) & 1.48 & 19 & 70 & 40.1 & 723 & 0.48 (143) & 17 & 1.4(-3) & 5.9(-5) & 2.8(4) \\
& 2.70 & 160 & 0.3 & 0.17 (49) & 24 & 80 & 100.0 & 98 & 37 & 2.2(-4) & 9.9(4) & 1.49 & 19 & 49 & 73.3 & 723 & 0.48 (143) & 17 & 1.4(-3) & 5.9(-5) & 2.8(4) \\
& 3.04 & 120 & 0.3 & 0.14 (42) & 12 & 48 & 73.7 & 93 & 24 & 1.6(-4) & 3.6(4) & 1.50 & 17 & 63 & 26.5 & 786 & 0.48 (143) & 15 & 1.5(-3) & 4.2(-6) & 2.6(4) \\
\hline\noalign{\smallskip}
NGC 7538
& 0.15 & 400 & 0.1 & 0.47 (36) & 16 & 22 & 23.2 & 364 & 17 & 1.1(-4) & 3.8(4) & 0.36 & 18 & 18 & 36.2 & 635 & 0.48 (38) & 10 & 1.2(-3) & 1.4(-7) & 2.3(4) \\
IRS9 & 0.19 & 320 & 0.1 & 0.42 (32) & 16 & 39 & 2.0 & 281 & 19 & 1.1(-4) & 3.7(4) & 0.37 & 13 & 18 & 39.8 & 615 & 0.44 (34) & 13 & 1.1(-3) & 9.0(-6) & 2.2(4) \\
$d$ = 2.65 kpc & 0.35 & 240 & 0.1 & 0.36 (28) & 24 & 39 & 52.5 & 171 & 33 & 1.1(-4) & 8.2(4) & 0.37 & 13 & 18 & 37.4 & 622 & 0.48 (38) & 11 & 9.9(-4) & 4.3(-6) & 2.2(4) \\
$R_{\rm ap}$ = 25.6\arcsec & 0.47 & 480 & 0.1 & 0.51 (40) & 16 & 22 & 17.2 & 440 & 15 & 1.2(-4) & 3.8(4) & 0.38 & 16 & 18 & 44.1 & 582 & 0.48 (38) & 18 & 1.1(-3) & 1.3(-7) & 2.3(4) \\
& 0.54 & 60 & 3.2 & 0.03 (2) & 12 & 34 & 22.2 & 38 & 27 & 7.6(-4) & 5.0(4) & 0.40 & 16 & 18 & 46.5 & 592 & 0.48 (38) & 22 & 1.1(-3) & 7.5(-7) & 2.6(4) \\
\enddata
\end{deluxetable*}
\end{sidewaystable*}

The parameters of the best-fit ZT models are listed in the left side
of Table~\ref{tab:models}. For each source, the best five models are
shown, ordered from best to worst as measured by $\chi^2$.  Note that
these are distinct physical models with differing values of $M_c$,
$\Sigma_{\rm cl}$ and/or $m_*$, i.e., we do not display simple
variations of $\theta_{\rm view}$ or $A_V$ for each of these different
physical models. Recall that the models are based on the Turbulent
Core Accretion theory (MT03), which links protostellar accretion rate
to core mass, clump mass surface density, and evolutionary stage (i.e.,
the mass of the protostar, $m_*$). Also the core radius, $R_c$, is
specified by $M_c$ and $\Sigma_{\rm cl}$. The accretion disk is always
assumed to have a mass that is one-third of $m_*$. These models (and those
of Robitaille et al., discussed below) are all for a single protostar
within a core. Note that even if observed cores are shown to contain
multiple sources, this approximation may still be reasonable if the
primary source dominates the luminosity of the system.

In general, the best-fit models have protostellar masses $m_*\sim
10$--50$\:M_\odot$ accreting at rates of
$\sim1\times10^{-4}$--$1\times10^{-3}\:M_\odot\:{\rm yr}^{-1}$ inside
cores of initial masses $M_c\sim 30$--500$\:M_\odot$ embedded in
clumps with mass surface densities $\Sigma_{\rm cl}\sim0.1$--3$\:{\rm
  g\:cm^{-2}}$ (note that this is the full range of $\Sigma_{\rm cl}$
covered by the model grid).

In many sources, the best five models have similar values of $\chi^2$,
i.e., they are of similar goodness of fit. In these cases, among the
best five models, there can also still be a significant variation in
model parameters, which illustrates degeneracies that exist in trying
to constrain protostellar properties from only their MIR to FIR
SEDs. There are also likely to be other models beyond the best five
that are still reasonable fits to the SEDs. However, we will not
explore these here, since already the consideration of just the best
five models shows the merits and limitations of this SED fitting.

Some of the degeneracies may be broken by using additional
information. One simple check is whether the size of the protostellar
core fits inside the aperture used to define the SED. The most
self-consistent situation is when $R_c$ is similar to $R_{\rm ap}$. If
$R_c\ll R_{\rm ap}$, then the peak of the SED is still likely to be
well-measured, but the long wavelength emission from cooler material
will be overestimated, i.e., the clump background subtraction would
have been underestimated. If $R_c\gg R_{\rm ap}$, then the observed
and model SED comparison is not self-consistent, although the peak of
the SED from the warmer material may still be contained in the
aperture. Better constraints come from using more detailed
morphological information, e.g.,
MIR/FIR intensity profiles along the outflow axis (Zhang et
al. 2013). A joint fitting of SEDs and image morphologies will be
carried out in a future paper in this series, following the methods of
Zhang et al. (2013).  Also, associated predictions of radio continuum
free-free emission (Tanaka et al. 2016) and observations of the mass
of the protostellar envelope are expected to be able to break
degeneracies in the models, and will be investigated in future works.


We now describe the results of the ZT SED model fitting of each of the
sources, using the best five models as examples. 

{\it AFGL~4029:} The best-fit model, with $\chi^2=1.08$, has
$m_*=12\:M_\odot$ accreting at $1.9\times10^{-4}\:M_\odot\:{\rm
  yr}^{-1}$ from a 30$\:M_\odot$ core in a $\Sigma_{\rm
  cl}=1.0\:{\rm g\:cm^{-2}}$ clump.
Such a source has $L_{\rm bol}=4.1\times10^{4}\:L_\odot$. However, an
almost equally good model (the next best fit with $\chi^2=1.32$) has a
protostar with $48\:M_\odot$ forming from a $160\:M_\odot$ core in a
$\Sigma_{\rm cl}=0.3\:{\rm g\:cm^{-2}}$ clump, seen nearly edge-on
with a much larger foreground extinction and with an almost $10\times$
larger bolometric luminosity. This specific example illustrates the
kinds of degeneracies that are present in fitting protostellar models
from MIR to FIR SEDs alone, and in this case such fitting is not very
constraining on the protostellar properties. However, we note that if
the source aperture size information is taken into account, then the
second case of a more massive, lower-density core has a radius of
17\arcsec\ that is significantly larger than the aperture radius used
to define the SED (11.2\arcsec). Thus, flux profile fitting may be
helpful here to break model degeneracies, in particular excluding
models that are too large for consistency with the size of the region
used to define the SED.


{\it AFGL437:} This source also has a best-fit model with
$\chi^2\simeq1$. However, here the fifth best model is a significantly
worse fit with $\chi^2\simeq2.2$. Most models involve fairly massive,
$\sim200\:M_\odot$ cores in a low $\Sigma=0.1\:{\rm g\:cm}^{-2}$
environment, but one example has a $50\:M_\odot$ core in a much higher
$\Sigma=3.2\:{\rm g\:cm}^{-2}$ clump. This case also has a viewing
angle that is close to the outflow cavity-opening angle so that there
are high levels of shorter wavelength emission, clearly
distinguishable among the SEDs.

{\it IRAS~07299:} Note that for this source, which lacks {\it
  Herschel} data, there are only four effective data points (plus the 3
to 8~$\rm \mu m$ data treated as upper limits) constraining the
models. The values of $\chi^2$ are small, i.e., about 0.5, for the
best-fit case. These models indicate $8$ to $16\:M_\odot$ protostars
in relatively low-$\Sigma$ cores viewed nearly edge-on are
preferred. However, these models have core radii that can be several
times larger than the aperture radius (but note that the 10 to 40~$\rm
\mu m$ emission in these models is generally quite concentrated in the
inner region of the core; ZTH14). Longer wavelength data would obviously be
helpful here to break some of these degeneracies.

{\it G35.20-0.74:} Here, the analysis also yields several models with a
similar goodness of fit, but now with relatively high values of
$\chi^2\simeq 2.6$. Inspection of the SEDs shows that the models
struggle to match the longer wavelength fluxes, i.e., $\geq160\:{\rm
  \mu m}$, with the model fluxes too small by about a factor of 1.7 at
this wavelength. These long wavelength data are sensitive to the
presence of cooler material. It is possible that better model fits
could be achieved if either background subtraction has been
underestimated and/or if the defined aperture radius is too large
and includes too much surrounding clump material. Other possibilities
are that for this source the approximation of there being a single
dominant source of luminosity is not as valid as in other cases, which
makes the model fits to be of generally poorer quality. As discussed
in \S\ref{S:G35.2}, there is evidence for G35.20-0.74 hosting a binary
system. Considering the results of the model fitting, there appears to
be a dichotomy amongst the best models, with higher-mass cores in
low-$\Sigma_{\rm cl}$ clumps and lower-mass cores (but still
$\sim100\:M_\odot$) in high-$\Sigma_{\rm cl}$ clumps giving similar
values of $\chi^2$. However, a more intermediate case of a
$200\:M_\odot$ core in a $\Sigma_{\rm cl}=0.3\:{\rm g\:cm^{-2}}$ clump
environment is also possible. Again, these results illustrate the
types of degeneracies that are present when trying to constrain
protostellar properties from such SED fitting.
%
We note that most of these models are in a relatively early stage of
formation, so the opening angles of their outflow cavities are quite
narrow, i.e., $\sim$20$^\circ$, and such angles are quite similar to
those implied by the morphologies shown by the high-resolution 11 and
18$\:{\rm \mu m}$ images of the source presented by De Buizer (2006);
see also Zhang et al. (2013b).


{\it G45.47+0.05:} For this source, the best-fit model has $\chi^2
\simeq 1.2$, with the fifth best model having a value of 1.7. The best
two preferred models have a similar goodness of fit and involve a 200
to 300$\:M_\odot$ core in a high-$\Sigma_{\rm cl}$ clump with a
current protostellar mass of $\sim30$ to $50\:M_\odot$ viewed nearly
edge-on. More intermediate viewing angles are returned for the next
best models (and also recall we have not fully explored the full range
of viewing angles that allowed for a given physical model).

{\it IRAS 20126:} Here, the best-fit value of $\chi^2 \simeq 2$. The
models prefer a lower $\Sigma_{\rm cl}\lesssim 0.3\:{\rm g\:cm^{-2}}$
clump environment with a $\sim$100 to $200\:M_\odot$ core that has
formed a protostar with $m_*\sim$12 to $24\:M_\odot$ viewed at
relatively large angles with respect to the outflow axis, i.e., with
the line of sight passing through the bulk of the core infall
envelope. Given the results of Chen et al. (2016), which favor a
$12\:M_\odot$ protostar from the kinematics of CH$_3$CN that may be
tracing the accretion disk, the third and fourth best-fit models may
be the most applicable in this case.

{\it Cep A:} This protostar also has best-fit models with $\chi^2
\simeq 2.2$, which rise to about 3 by the fifth model. The best two
models prefer $\Sigma_{\rm cl}=0.3\:{\rm g\:cm^{-2}}$ with
$M_c=160\:M_\odot$ and $m_*=16$ or 12$\:M_\odot$ viewed at angles of
44$^\circ$ to $29^\circ$ with about 100~mag of foreground extinction.

{\it NGC~7538 IRS9:} This source has an SED that is very well fit by
the ZT RT models, with $\chi^2 \simeq 0.15$ for the best case, rising
to 0.5 for the fifth best physical model. Most of the models prefer
$\Sigma_{\rm cl}=0.1\:{\rm g\:cm^{-2}}$. The best-fit model has a
$400\:M_\odot$ core with a $16\:M_\odot$ protostar, viewed at a
relatively small angle with respect to the outflow axis, i.e., the
line of sight passes close to the ouflow cavity boundary, avoiding
most of the infall envelope. The SEDs of such models are relatively
flat from $\sim$20 to $\sim 100\:{\rm \mu m}$. We note that the small
viewing angle of $22^\circ$ of the best-fit model is similar to the
value of $\sim20^\circ$ inferred from the HCO$^+$ outflow by Sandell
et al. (2005; see \S\ref{S:NGC7539IRS9}).


\subsubsection{Robitaille et al. Model Fitting Results}

\begin{figure*}
\epsscale{1.00} \plotone{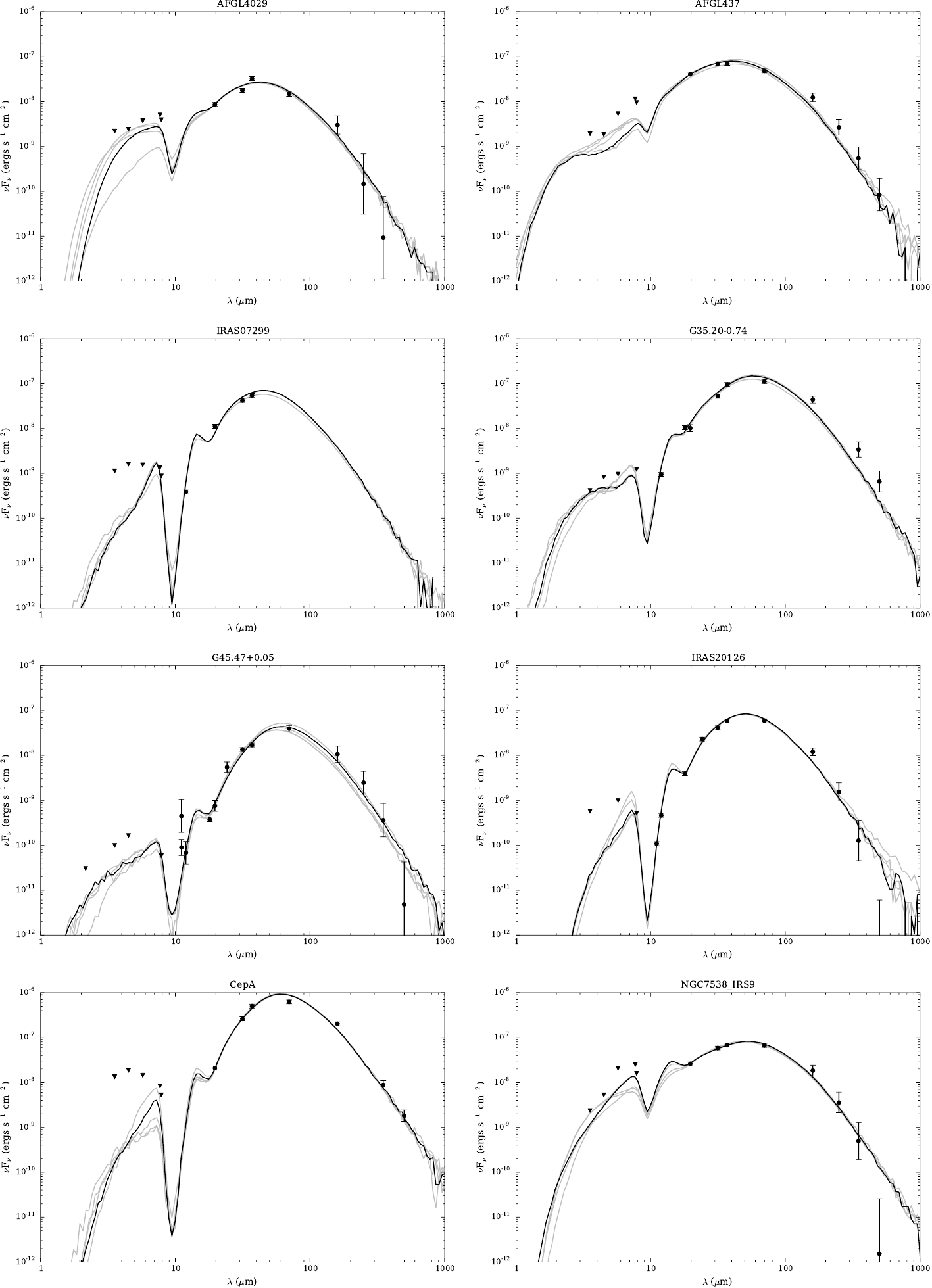}
\caption{
Protostar model fitting to the fixed aperture, background-subtracted
SED data using the Robitaille et al. (2007) model grid. For each
source, the best-fit model is shown with a solid black line and the
next four best models are shown with solid gray lines. Flux values are
those from Table~\ref{tab:flux}. Note that the data at
$\lesssim8\:{\rm \mu m}$ are treated as upper limits (see text). Also,
the fitting method sets the data point to be at the middle of the
error bar range. The resulting model parameter results are listed in
Table~\ref{tab:models}.
\label{fig:SEDsRob}}
\end{figure*}


In Figure~\ref{fig:SEDsRob}, we show the results of fitting the
Robitaille et al. (2007) models to the fiducial SEDs. The parameters
of the best five models are also shown in the right side of
Table~\ref{tab:models}. The values of $\chi^2$ for the Robitaille et
al. models are quite similar to those of the ZT models, with a modest
tendency to return slightly poorer fits, even though the sampling in
some parameters, like $m_*$, is finer than the ZT model grid and the ZT
models involve fewer free parameters.

With the Robitaille et al. models, a common occurrence is that the disk
accretion rates are much lower than in the ZT models, often
$\sim100\times$ smaller (and occasionally $\sim10^3\times$
smaller). In some cases, the models do not require any disk component
(indicated by ``...'' in the tabulated accretion rates). The envelope
infall rate is always much larger than the disk accretion rate, so the
models are not physically self-consistent, at least in the context of
having a steadily accreting system. Lower disk accretion rates mean a
smaller bolometric luminosity and so to compensate the Robitaille et
al. results can involve larger protostellar masses than the ZT models,
e.g., in the cases of IRAS~07299, G35.20-0.74, Cep A, or smaller
overall extinctions due to lower column density cores and/or more
face-on viewing angles.

The outer core envelope radii, $R_{\rm env}$, are also typically quite
large, i.e., $\sim$0.5~pc. Only for the distant source G45.47+0.05 are
these smaller than the aperture size used to define the SED. In the
other sources, $R_{\rm env}>R_{\rm ap}$, sometimes by factors of five
or so.
Thus, most of these models are not internally self-consistent with the
observations.

Considering the particular case of G35.20-0.74 is instructive. As with
the ZT models, the best-fit Robitaille et al. models underpredict at
long wavelengths and (slightly) overpredict near the peak of the
SED. A protostellar mass of $m_*=20\:M_\odot$ is estimated, but with a
disk accretion rate of only $2.8\times10^{-7}\:M_\odot\:{\rm yr}^{-1}$
(so accretion power is negligible). On the other hand, the envelope
mass infall rate is $1.6\times 10^{-3}\:M_\odot\:{\rm yr}^{-1}$. The
viewing angle is found to be 87$^\circ$, so that the outflow axis
would be close to the plane of the sky, which is very different from
the result of the best ZT model. Such a geometry would not be expected
to lead to strong asymmetries in the MIR/FIR morphologies of the blue-
and redshifted outflow cavities.

As discussed below, future studies that use additional constraints
such as observations of the radio continuum flux (which is sensitive
to protostellar mass), MIR-FIR image intensity profiles along and
transverse to the outflow cavity axis (which help measure the density
and temperature structure of the core infall envelope and outflow),
along with other tracers of the gas content, can help test between the
relative validity among the Robitaille et al. models and in comparison
to the ZT models.

\subsection{Discussion}

The above considerations illustrate current capabilities, including
difficulties and uncertainties, of determining protostellar properties
from simple SED fitting methods. We consider the results of the ZT
model fitting to be more reliable since the models are designed with
the typical expected properties of massive protostars in mind and they
yield results that are internally self-consistent both physically
(i.e., accretion rates through the disk are directly related to infall
rates in the core envelopes; such high disk accretion rates are likely
to be needed to drive powerful outflows) and observationally (i.e.,
the cores are more compact and are generally a better match to the
aperture sizes used to define the SEDs).

Future work can help test the models further. For example, the
estimated disk accretion rates can be compared with observed mass
outflow rates to see if they are consistent with theoretical models of
disk winds and/or X-winds (e.g., Caratti o Garatti et al. 2015;
Beltr\'an \& de Wit 2016). The model and observed image intensity
profiles along and transverse to the outflow axis can be compared to
better constrain the outflow opening angle and orientation (i.e.,
$\theta_{\rm view}$; e.g., Zhang et al. 2013). The latter can also be
compared to that estimated from a study of the kinematics of the
outflowing gas. Predictions for internal density and temperature
structures within the core can be tested with higher angular
resolution observations, e.g., of MIR emission (e.g., Boley et
al. 2013), of sub-millimeter/millimeter dust continuum (e.g., Beuther
et al. 2013), and with specific temperature diagnostics such as $\rm
NH_3$ inversion transitions (e.g., Wang et al. 2012). Kinematics of
core envelope infall (e.g., Wyrowski et al. 2016) and disk rotation
(e.g., S{\'a}nchez-Monge et al. 2013) can also be probed with
molecular lines.  Magnetic field structures around the protostars can
be inferred from observations of polarized dust continuum emission
(e.g., Girart et al. 2009; Zhang et al. 2014b). The density and
temperature structures of the radiative transfer models, along with
their evolutionary histories, provide a framework for developing and
testing astrochemical models of core envelopes, disks, and outflows
(e.g., Doty et al. 2006; Drozdovskaya et al. 2014; Zhang \& Tan
2015). Radio continuum emission tracing, e.g., photoionized gas, can be
searched for (e.g., Rosero et al. 2016) and compared to theoretical
predictions (e.g., Tanaka et al. 2016).





\begin{figure*}
\epsscale{1.10}
\plotone{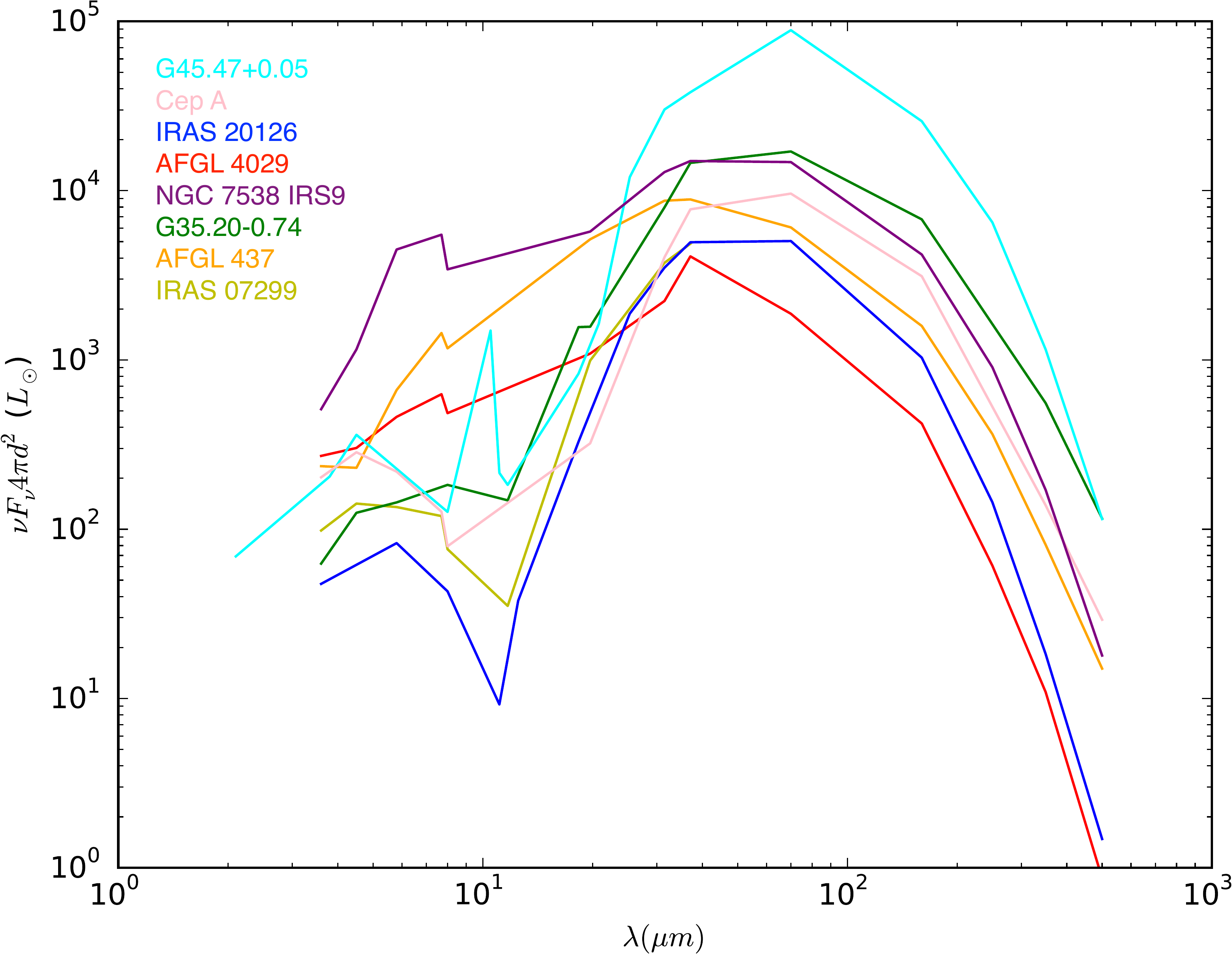}
\caption{
Bolometric flux-weighted SEDs of the eight SOMA protostars analyzed in
this paper. The ordering of the legend is from high to low ZT best-fit
model luminosity (top to bottom).}\label{fig:Lbol}
\end{figure*}

By eventually studying a large sample of protostars spanning a range
of environments, masses, and evolutionary stages, we hope to discern
general trends in star formation activity. For example, do protostars
in higher $\Sigma_{\rm cl}$ clump environments have higher accretion
rates, as would be predicted by the Turbulent Core Model? Or do such
environments involve protostars forming with different accretion
mechanisms? Are there systematic trends in SED shape with clump
environment, core mass, protostellar mass or luminosity?

As a first step in such directions, in Figure~\ref{fig:Lbol} we show
the bolometric luminosity SEDs of the eight protostars, i.e., the $\nu
F_\nu$ SEDs have been scaled by $4\pi d^2$, so that the height of the
curves gives an indication of the luminosity of the sources, assuming
isotropic emission. This figure allows one to visualize the range in
luminosities present in the sample, along with any potential trends in
SED shape. However, on inspecting the distributions, we do not
perceive any obvious trends in SED shape with luminosity, although
this is perhaps not so surprising given the current sample size.


We can compare the ordering of the vertical height of these
distributions with the rank ordering of the predicted true luminosity
of the protostars from the best-fit ZT models (the legend in
Fig.~\ref{fig:Lbol} lists the sources in order of decreasing ZT model
luminosity).
There is some, but not perfect, correspondence with the flux ordering
seen in the figure. Differences are most likely due to varying levels
of foreground extinction, local extinction in the core envelope (e.g.,
AFGL 4029's formal best-fit ZT model has a relatively low envelope
mass and wide outflow cavity, so a large fraction of its luminosity
would not be re-radiated in the MIR to FIR) and anisotropic beaming
(i.e., the ``flashlight effect,'' Yorke \& Bodenheimer 1999). The
latter effect likely boosts NGC 7538 IRS9's apparent bolometric
luminosity SED compared to that expected based on its intrinsic
bolometric luminosity. Such non-intrinsic effects illustrate the need
for larger samples of protostars, i.e., eventually, statistically
significant samples will be required as a function of environment,
mass, and evolutionary stage. This is the eventual goal of the SOMA
Survey.

\section{Conclusions}

We have presented an overview and first results of the SOMA Star
Formation Survey. The survey's scientific rationale is to test
predictions of Core Accretion models of massive star formation,
specifically the MIR to FIR thermal dust emission, including the
influence of outflow cavities. We have presented results for the first
eight sources observed in the survey. These tend to show extended MIR
and FIR emission that aligns with known outflows, and being brighter
on the near-facing, blueshifted side, which are predictions of Core
Accretion models that involve high mass surface density cores. In
principle, unrelated foreground extinction could mimic these results,
but the consistency of the observed multiwavelength morphologies in
the sample provides strong support for the Core Accretion scenario.

Global SEDs have been constructed and the effects of choices of
aperture definition and background subtraction investigated. Our
fiducial method is an SED derived from a fixed aperture and including
an estimate of background subtraction, i.e., the emission from the
surrounding clump environment.

These SEDs have been used to constrain properties of the protostars by
comparing with theoretical radiative transfer models of massive star
formation via the Turbulent Core Accretion model. These yield
protostellar masses $m_*\sim 10$--50$\:M_\odot$ accreting at rates of
$\sim1\times10^{-4}$--$1\times10^{-3}\:M_\odot\:{\rm yr}^{-1}$ inside
cores of initial masses $M_c\sim 30$--500$\:M_\odot$ embedded in
clumps with mass surface densities $\Sigma_{\rm cl}\sim0.1$--3$\:{\rm
  g\:cm^{-2}}$. We note that these are results from using a model grid
with a relatively coarse sampling of initial core masses and clump
envelope mass surface densities, yet quite reasonable fits are found.
The derived accretion rates are comparable to the values estimated by
other means, e.g., via observed infall rates in core envelopes (e.g.,
Wyrowski et al. 2016) and via mass outflow rates (e.g., Beltr\'an \&
de Wit 2016). However, there can be significant degeneracies in the
parameters of models that provide good fits to the SEDs. Breaking
these degeneracies will require additional observational contraints,
such as using predictions of image intensity profiles (e.g., Zhang et
al. 2013) or radio continuum emission that traces ionized gas (e.g.,
Tanaka et al. 2016).

Comparison with the widely used Robitaille et al. (2007) model grid
finds large differences, especially in the derived disk accretion
rates. We suspect that these differences are due, at least in part, to
there being a wider choice of free parameters in the Robitaille et
al. grid, which can lead to models that we consider less physically
realistic, i.e., high mass infall rates in the core envelope but small
disk accretion rates.

Finally, we emphasize the importance that SOFIA-FORCAST
observations in the wavelength range $\sim 10$ to $40\:{\rm \mu m}$
have for constraining the theoretical models. In combination with {\it
  Herschel} 70 to 500$\:{\rm \mu m}$ data, they allow measurement of
the thermal emission that defines the peak of the SED and probes the
bulk of the bolometric flux. We consider this thermal emission simpler
to model than that at shorter wavelengths, $\lesssim 8\:{\rm \mu m}$,
which is more affected by emission from PAHs and transiently heated
small dust grains.

Future papers in this series will present additional sources,
especially probing a wider range of environmental conditions,
evolutionary stages, and protostellar core masses. Additional analysis
that examines and models flux profiles along outflow cavity axes will
be carried out, following methods developed by Zhang et
al. (2013b). Ancillary observations that trace the outflowing gas will
also be presented.

\acknowledgments We thank an anonymous referee for helpful comments,
which improved the manuscript. J.C.T. acknowledges several
NASA-USRA-SOFIA grants that supported this research. This work is
based in part on observations made with the NASA/DLR
Stratospheric Observatory for Infrared Astronomy
(SOFIA).
This work is also based in part on observations made with the
\textit{Spitzer Space Telescope}, which is operated by the Jet
Propulsion Laboratory, California Institute of Technology under a
contract with NASA. This work is additionally based on observations
obtained at the Gemini Observatory (program GS-2005B-Q-39),
which is operated by the Association of Universities for Research in
Astronomy, Inc., under a cooperative agreement with the NSF on behalf
of the Gemini partnership: the National Science Foundation (United
States), the National Research Council (Canada), CONICYT (Chile),
Ministerio de Ciencia, Tecnología e Innovación Productiva (Argentina),
and Ministério da Ciência, Tecnologia e Inovação (Brazil). The lead
author was also a Visiting Astronomer at the \textit{Infrared
  Telescope Facility}, which is operated by the University of Hawaii
under contract NNH14CK55B with the National Aeronautics and Space
Administration.


\facility{SOFIA (FORCAST); {\it Herschel} (PACS, SPIRE); {\it Spitzer} (IRAC); Gemini (T-ReCS, MICHELLE); IRTF (NSFCam, MIRLIN)}


\end{document}